\documentclass[preprint,12pt]{elsarticle}

\usepackage{amssymb}
\usepackage{svg}
\usepackage{amsmath}
\usepackage{booktabs}
\usepackage[hyphens]{url}
\usepackage{siunitx}
\usepackage{url}
\usepackage{adjustbox}
\usepackage[utf8]{inputenc}
\usepackage[english]{babel}
\usepackage{hyperref}
\usepackage{nomenclature_style}
\usepackage{nomenclature}
\usepackage{color}
\usepackage{xcolor}
\definecolor{darkblue}{rgb}{0,0.2,0.49}
\hypersetup{ colorlinks=true,
linkcolor=darkblue,
filecolor=darkblue, 
citecolor=darkblue,
urlcolor=darkblue 
}

\usepackage{placeins}
\usepackage{caption}
\usepackage{subcaption}
\usepackage{setspace}
\usepackage{cleveref}
\usepackage{todonotes}

\crefname{figure}{Figure}{Figures}
\crefname{table}{Table}{Tables}
\crefname{equation}{Equation}{Equations}
\crefname{appendix}{}{}
\crefname{section}{Section}{Sections}

\newcommand{\eg}{\textit{e.g.}}

\journal{Elsevier}

\newcommand{\periodlength}{L^{period}}
\newcommand{\seasScaleFac}{\alpha_s}
\newcommand{\geographicCapacity}{A^c_n}
\newcommand{\remainingInitCap}{\bar{x}^{a}_{n,i}}
\newcommand{\lifetime}{i^{life}_a}

\newcommand{\flowSource}{y^{c,source}_{n,p,h,i,\omega}}
\newcommand{\flowSink}{y^{c,sink}_{n,h,i,\omega}}
\newcommand{\flowOut}{y^{c,trans}_{n,m,h,i,\omega}}
\newcommand{\flowIn}{y^{c,trans}_{m,n,h,i,\omega}}
\newcommand{\commodityDemand}{D^c_{n,h,i,\omega}}
\newcommand{\lostLoad}{y^{ll}_{n,h,i,\omega}}

\newcommand{\investmentVariable}{x^a_{n,j}}
\newcommand{\capacityVariable}{v^a_{n,i}}

\newcommand{\operationsAsset}{y^a_{n,i,h,\omega}}

\newcommand{\capacityStorage}{v^{c,stor}_{n,i}}
\newcommand{\storageOperational}{w^{c}_{n,h,i,\omega}}
\newcommand{\storageOperationalPrev}{w^{c}_{n,h-1,i,\omega}}
\newcommand{\storageCharge}{y^{c,chrg}_{n,h,i,\omega}}
\newcommand{\storageDischarge}{y^{c,dischrg}_{n,h,i,\omega}}

\begin{document}

\begin{frontmatter}




\title{Decarbonizing the European energy system in the absence of Russian gas: Hydrogen uptake and carbon capture developments in the power, heat and industry sectors}



\author[inst1]{Goran Durakovic\corref{cor1}}
\cortext[cor1]{Corresponding author.}
\ead{goran.durakovic@ntnu.no}
\pdfstringdefDisableCommands{
  \def\corref#1{<#1>}
}

\author[inst1]{Hongyu Zhang}

\author[inst2]{Brage Rugstad Knudsen}

\author[inst1]{Asgeir Tomasgard}

\author[inst1]{Pedro Crespo del Granado}

\affiliation[inst1]{organization={Department of Industrial Economics and Technology Management},
            addressline={Norwegian University of Science and Technology}, 
            city={Trondheim},
            country={Norway}}

\affiliation[inst2]{organization={SINTEF Energy Research},
            addressline={Sem Sælands vei 11},
            city={Trondheim},
            country={Norway}}

\begin{abstract}

Hydrogen and carbon capture and storage are pivotal to decarbonize the European energy system in a broad range of pathway scenarios. Yet, their timely uptake in different sectors and distribution across countries are affected by supply options of renewable and fossil energy sources. 
Here, we analyze the decarbonization of the European energy system towards 2060, covering the power, heat, and industry sectors, and the change in use of hydrogen and carbon capture and storage in these sectors upon Europe's decoupling from Russian gas. 
The results indicate that the use of gas is significantly reduced in the power sector, instead being replaced by coal with carbon capture and storage, and with a further expansion of renewable generators. Coal coupled with carbon capture and storage is also used in the steel sector as an intermediary step when Russian gas is neglected, before being fully decarbonized with hydrogen. Hydrogen production mostly relies on natural gas with carbon capture and storage until natural gas is scarce and costly at which time green hydrogen production increases sharply. The disruption of Russian gas imports has significant consequences on the decarbonization pathways for Europe, with local energy sources and carbon capture and storage becoming even more important.

\end{abstract}



\end{frontmatter}




\section{Introduction} \label{sec:introduction}

In the wake of the disruption of Russian gas supply to Europe, European Union (EU) policymakers are reshaping incentives and measures to reduce dependency on Russian fossil fuels and maintain the pace of emission reduction and decarbonization efforts~\citep{RePowerEU}. Sector-specific and cross-sectorial plans are being rolled out to adapt implementation plans for decarbonization and electrification, promote necessary technology developments, and ensure the economic viability of transition with a sharpened competition for clean energy. Recently, the EU launched the Net-Zero Industry Act \citep{NetZeroEIndustryACT} as a part of the Green Deal Industrial Plan, promoting regulatory conditions that facilitate faster scale up of technologies that are crucial for sectors that must reach net-zero by 2050, such as wind and solar, renewable hydrogen and CO$_2$ storage. 

The disrupted Russian gas supplies and geopolitical instabilities increase energy scarcity in the European energy market and reinforce the price pressure and volatility for both fossil and renewable energy. Competition for clean energy increases, while limitations in the availability of rare and vital metals together with supply constraints create delays and cost challenges for several large-scale renewable energy projects. The prevailing energy crisis and rapidly evolving energy landscape in Europe present ambiguous energy transition trajectories, especially with sustained removal~\citep{Pedersen2022} of Russian gas supplies. A large share of hydrogen is a recurring scenario, \eg,~\cite{SECK2022112779}, yet to the best of our knowledge little studied upon disruption of Russian gas supply. Several European countries have reoriented to LNG imports, while the ambitions for penetration of hydrogen as a clean fuel are maintained~\citep{RePowerEU}. Current estimations for future hydrogen consumption appear to be at odds with emerging data~\cite{RikvanRossum2022EuropeanBackbone}. The impact of limited energy supplies on prioritization of and strategies for remaining possible decarbonization options should thus be lifted. \citet{Pedersen2022} addressed this topic, focusing  particularly on cross-sector distribution of capacities and use of renewable energy across sectors to adhere with the 1.5\textdegree C climate target. They showed that the 1.5\textdegree C target can be maintained without Russian gas supplies, while a 2\textdegree C target is greater affected.  \citet{Mannhardt2023} explored the effects of collective demand reduction across sectors as a response to disrupted Russian gas supply, with the objective of reducing energy consumption.  \citet{Klaaßen2023} used a meta-analytical approach to explore shifts in needed power and transport investments to maintain climate targets as a consequence of Russian gas removal in EU.

This paper broadens the impact analysis of persistent removal of Russian gas supply, focusing particularly on the uptake hydrogen and CCS in the power, heat and industry sectors. To this end, endogenous hydrogen demand modelling in energy system modes is needed to achieve more accurate projections. Such an integration has so far been overlooked in the scholarly discourse. 
The open-source power-system model EMPIRE model \cite{Backe2022} is applied and its scope is extended by enhancing its analytical capability to scrutinise the role of natural gas and hydrogen in the prospective European energy infrastructure. Originally designed for long-term European power system expansion planning, the EMPIRE model has since been augmented to encapsulate CCS~\cite{Turgut2021AnEurope}, domestic heating systems~\cite{Backe2022} and hydrogen production~\cite{Durakovic2023PoweringPrices,Durakovic2023AreAnalysis}. Using EMPIRE, our focus rests on the modelling of hydrogen production technologies, which include electrolyzer and natural gas reforming processes both with and without CCS, while considering scarcity of both electricity and natural gas. Furthermore, we evaluate energy consumption and the feedstock requirements of major industry sectors, such as cement, steel, ammonia, and refinery. The modelling approach for the power and heat sectors is informed by~\cite{Backe2022}, while the energy consumption figures for the transport sector are derived from external references. Our methodological approach seeks to illuminate the fuel and feedstock switch from natural gas to hydrogen within the future European energy system. To maintain the tractability of the model, we employ linear programming while retaining existing features of the EMPIRE model, including the handling of short-term uncertainty and multi-period investment planning. 

The main contributions in this paper include: (1) the incorporation of endogenous hydrogen demand within a large-scale, long-term energy system investment model, (2) detailed modelling of the energy consumption and feedstock demand in key industry sectors, and (3) a comprehensive analysis of the influence of natural gas price and availability on hydrogen production, and the subsequent decarbonization implications for the power, heat, and industry sectors in Europe.

The structure of the paper is as follows: Section \ref{sec:literature review} furnishes background information concerning the industry sector's role in energy systems, the prospective impact of hydrogen, and the use of CCS. Section \ref{sec:methodology} elucidates the adopted methodology and data sources. Section \ref{sec:results} presents and interprets our computational results. Finally, Section \ref{sec:conclusion} provides concluding thoughts and directions for future research.

\section{Literature review}
\label{sec:literature review}

 
In the following, we present a brief overview of relevant literature on the energy consumption and decarbonization of the industry sectors and its representation in energy system planning models, demand side flexibility in industry sectors, and the potential role of CCS and hydrogen in the industry sector. 

\subsection{The industry sector in the energy system}
In 2021, the industry sector accounted for 25.6\% of the final energy consumption in the EU~\cite{EuropeanEnvironmentAgencyEEAAgency}. It was the third largest energy consumer among all sectors. Also, the industry sector accounted for 22\% total emissions in the EU with 757 million tonnes CO$_2$. Therefore, it is important to decarbonize this sector. An optimisation model for the simulation and operational optimisation of the industry sector with a high level of detail was developed~\cite{Wiese2018ConceptualDenmark}. This bottom-up model was demonstrated in the Danish energy system. The pathway of the energy transition is simulated but not optimised. Although these models include sufficient operational details, the optimal investment for the transition in the industry sector was not investigated. In this paper, we aim to fill this gap by including the investment planning of the industry sector in a long-term stochastic energy system planning model. We focus on modelling the energy consumption of the industry sector, including cement, steel, ammonia and refinery. These are the major energy consumers in the industry sector. In the following, we present the background knowledge on modelling production processes in these sectors.

Cement production usually consists of raw materials handling, pyroprocessing, milling and bagging~\cite{EuropeanEnvironmentAgency20192.A.1Agency}. The CO$_2$ emits during the pyroprocessing phase, where the raw materials mix needs to be heated up to produce clinker. The detailed cement production processes were provided by~\citet{Alsop2019TheEDITION}. Traditionally, the fuel used to generate heat is natural gas. Different CCS technologies in the cement industry were reviewed~\cite{Hills2016CarbonRetrofitting}. The use of hydrogen in cement production is a relatively new area of study and represents an interesting pathway for decarbonizing the cement industry. A techno-economic assessment of using by-product oxygen from water electrolysis in hydrogen production for CCS in clinker production demonstrates potential cost advantages and highlights considerations around supply reliability and transport distance~\cite{Nhuchhen2022DecarbonizationEconomy}. The papers included detailed cement production processes but were limited to the cement industry only. In this paper, we consider both hydrogen fuel switch and CCS, and include the decarbonization of the cement industry in a large energy system planning problem.

The steel production process involves the extraction of iron from its ore, purification, and conversion into steel, typically through the blast furnace-basic oxygen furnace method or the electric arc furnace method. Steel-making and continuous casting is usually the bottleneck in iron and steel production. An integer programming model was developed to optimise this process by~\citet{Tang2002Steel-makingRelaxation}. A techno-economic model was developed for evaluating four alternative primary steelmaking routes~\cite{Fischedick2014Techno-economicTechnologies}. The authors investigated the economic and technical viability of innovative primary steel production methods in Germany until 2100 by comparing three new ore-based steelmaking routes to the traditional blast furnace method. The study showed that with rising prices for coal and CO$_2$ allowances, blast furnace-based routes might become unprofitable, making hydrogen direct reduction and iron ore electrolysis economically attractive due to higher energy and raw material efficiency together with the potential to meet 80\% reduction targets. However, high investment costs and electricity price dependency could hinder profitable implementation without further subsidies before 2030–2040.

Traditionally, natural gas is used as a feedstock for ammonia production plants to produce hydrogen locally via steam reforming and then produce ammonia from hydrogen~\cite{Simonelli2014FRAMEWORKINDUSTRY-AMMONIA}. Optimisation models for the production optimisation of chemicals can be hard to solve due to the inclusion of complex constraints. A trust region filter method for the black-box optimisation problem was proposed and was applied to solve an ammonia synthesis problem~\cite{Eason2016AOptimization}. Here, due to the problem size and research focus, we simplified the modelling of ammonia production. In addition, we consider ammonia production from the purchased hydrogen from a hydrogen system. 

Hydrogen is used to reduce the sulfur content of diesel fuel in the refinery industry. Traditionally, hydrogen is produced on-site with some emissions. A single objective optimisation model is proposed to maximise hydrogen production in an oil refinery at steady state condition~\cite{Sarkarzadeh2019ModelingRefinery}. The study showed that the main advantages of the optimized process were the higher hydrogen production at lower steam capacity in the plant and higher hydrogen production in reforming and shifting reactors. A linear programming model was developed to optimise the hydrogen distribution network for the refinery industry, and an efficient network design has been achieved with a 30\% reduction in hydrogen utility usage~\cite{Fonseca2008HydrogenStudy}. Most of the literature considered the optimisation of hydrogen production on-site, and the emissions from producing hydrogen were not sufficiently addressed. In this paper, we combine the refinery sector with other industry sectors and consider acquiring hydrogen from the system for the refinery processes. 

The energy consumption of the industrial sector is a large share of the total energy consumption. However, in most of the existing energy system investment planning models, the industry is modelled simplistically. In~\cite{Backe2022}, the energy consumption is only modelled exogenously, and the energy transition in such a sector is not sufficiently modelled. In this paper, we aim to fill this research gap by including sufficiently detailed operational modelling of the industry sectors in a long-term energy system planning model and analyse the energy transition in the industry sector. 

Due to the higher penetration of uncontrollable renewable energies, demand-side management has become an increasingly interesting and important topic. It is important to harness renewable energies better. There is a potential in the industry sector to shift their production activities according to energy availability. \citet{Zhang2016Enterprise-widePerspectives}  pointed out that the active management of electricity demand by power-intensive process industries is an important part of demand side management. A comprehensive review of the existing works on enterprise-wide optimisation for industrial demand side management was presented. As a major energy consumer, demand-side management in steel plants can help stabilise the power grid~\cite{Castro2020IndustrialReplacement}. The authors developed a new mixed integer linear programming model to optimize electric arc furnace operations in steel plants, showing that despite low electricity prices, high-power modes are largely avoided due to their less energy-efficient nature and higher electrode consumption, emphasizing the importance of electrode replacement in reducing overall costs. In this paper, we include industrial demand-side flexibility, which can be a significant source of flexibility~\cite{Gils2014AssessmentEurope}, by allowing each industry sector to shift their production by some percentage of their capacities. 

\subsection{Hydrogen in energy systems}
From the literature above, we can see that hydrogen can be used in multiple industry sectors, and in this paper, we systematically model the potential hydrogen demand in industry. In addition to providing fuel and feedstock for industrial production processes, hydrogen as a clean energy carrier can be used in other sectors, such as power and heat and can be important for the energy transition in general. We provide some literature on hydrogen in the energy transition in the following. 

Cloete et. al~\cite{Cloete2022BlueWorld} investigated the potential trade channels for energy exporters in a low-carbon future using a new model. They found that natural gas imports with CO$_2$ capture is the least costly solution. However, exporting blue hydrogen or steel produced via hydrogen reduces CO$_2$ handling and is a viable diversification strategy for fossil fuel exporters like Norway, despite moderately higher costs. \citet{Moreno-Benito2017TowardsDevelopment} extended the SHIPMod optimization framework to develop a sustainable hydrogen infrastructure for the UK's transition towards a low-carbon transport system. The extended model includes economies of scale, road and pipeline transportation, and CCS technologies. The authors found that the most cost-effective hydrogen production method that maintains low carbon emissions is natural gas reforming with CCS. Bødal et. al ~\cite{Bdal2020DecarbonizationStudy} proposed a cost-minimizing model to optimize investments in electricity and hydrogen infrastructure under various low-carbon scenarios. They found that in Texas, by 2050, hydrogen produced from both electricity and natural gas is cost-effective for emissions reduction, offering system flexibility and enabling high renewable energy shares with less battery storage. However, the results showed that the shift from electrolysis to steam methane reforming for hydrogen production depends on carbon pricing and hydrogen demand. A mixed-integer linear programming model was proposed to use offshore energy hubs to produce and store green hydrogen offshore for the decarbonization of the Norwegian continental shelf~\cite{zhang2022_OEHb} and the European energy system~\cite{Zhang2022OffshoreShelf}. The REORIENT model was proposed to integrate investment, retrofit and abandonment planning in a single stochastic mixed-integer linear programming for the long-term planning of the European energy system~\cite{Zhang2023IntegratedDecomposition}. The results showed that the REORIENT model could yield 24\% lower investment cost in the North Sea region than the traditional investment-planning-only model. 

Only a few published studies have explored the integrated natural gas, CCS and hydrogen value chains in multi energy system models.  \citet{Sunny2020} developed a H2–CCS value-chain modelling framework as a resource task network, incorporating  the specification of exogenous demand that can be satisfied using hydrogen and other alternatives. Hydrogen and CCS infrastructure was optimized, yet few details on the demand side, particularly the industry sector, were included, and the power sector was omitted in the model. \citet{Reigstad2022} analyzed future hydrogen demand and infrastructure for hydrogen production, transport and storage with a specific focus on Germany, the UK, the Netherlands, Switzerland and Norway. The analysis also included the use of hydrogen and its combination with CCS for decarbonization of both industry and transport, still with exogenous demand. The studies of \cite{Pedersen2022} and \cite{Victoria2022} applied the PyPSA-Eur-Sec model including options to invest in hydrogen production using steam methane reforming with or without CCS and electrolysis. Options for autothermal reforming with CCS, constituting improved efficiency and reduced CO$_2$ emissions were not included in the model. Resorting to a deterministic approach, stochasticity in renewable generation was omitted in the model and a 3 hour time resolution was used, thereby limiting the impact of variability in renewable generation in their analysis. \citet{SECK2022112779} analyzed the potential of low-carbon and renewable hydrogen in decarbonizing the European energy system according to the set EU targets, using a three-level, deterministic modelling approach with a detailed European TIMES-type model (MIRET-EU), an aggregated model for the European energy system, and a dedicated model for assessing hydrogen import options for Europe (HyPE). An emerging feature of this approach was the ability of endogenous cost reductions based on technology deployment in the model.

\begin{table}[t]
    \centering
        \caption{Comparison of this paper with relevant literature.}
    \label{tab:literaturecomparison}
    \resizebox{\columnwidth}{!}{
    \begin{tabular}{lccccccccc}
    \toprule
        \textbf{Ref.} & \textbf{Optimization} & \textbf{Multi-period} & \textbf{Stochastic} & \textbf{Power} & \textbf{Heat} & \textbf{Industry} &\textbf{Hydrogen} & \textbf{CCS\footnotemark} & 
        \textbf{Natural gas}\footnotemark \\
        \cite{SECK2022112779}&X &X & &X &X  &X  &X& X&\\
        \cite{Sunny2020}&X &X &  & &X  &X  &X& X&\\
        \cite{Pedersen2022}&X &X & & X &X  &X  &X& X&\\
        \cite{Backe2022} &X &X &X &X &X & \\
        \cite{Zhang2022OffshoreShelf} &X &X &X &X & X & &X& \\
        \cite{Bdal2020DecarbonizationStudy}  &X & & &X &  & &X& \\
        \cite{Fischedick2014Techno-economicTechnologies}  & & & & &  &X (only steel) &X & \\
        \cite{Nhuchhen2022DecarbonizationEconomy}& & & & &  &X (only cement) &X & \\
        \cite{Fonseca2008HydrogenStudy}&X & & & &  &X (only refinery) &X & \\
        This paper & X & X & X & X & X & X &X & X  & X\\
        \bottomrule
    \end{tabular}
    }
\end{table} 
\footnotetext[1]{This column marks the papers that include the development of the CCS transport chain, as well as the sequestration of CO$_2$.}
\footnotetext[2]{The natural gas column designates those papers that model the natural gas reserves and the production from these, or import from LNG terminals, along with transport through the natural gas pipeline network.}

A comparison of this paper with relevant literature is presented in \cref{tab:literaturecomparison}. In addition, for a more detailed literature review on hydrogen in energy systems, we refer the readers to \citet{Agnolucci2013DesigningScales}, who reviewed hydrogen literature across different spatial scales, and \citet{Li2019c}, who reviewed optimization literature on hydrogen supply chains.


\section{Methodology and data} \label{sec:methodology}

EMPIRE~\cite{Skar2016b,Backe2022} is used in this paper, formulated as a multi-horizon~\cite{Kaut2014b} stochastic~\cite{Birge2011b} mathematical problem. EMPIRE minimizes the investment and operational costs for power production, transmission, and storage. While EMPIRE was originally a power sector model, it has since been expanded considerably with an explicit model for the domestic heating demand~\cite{Backe2023ExploringNeighbourhoods}, and also the production of green~\cite{Durakovic2023PoweringPrices} and blue~\cite{Durakovic2023AreAnalysis} hydrogen to meet an exogenous demand. In this work, EMPIRE has been expanded to include the option to develop a CCS chain, and it now includes the industry sector together with the hydrogen sector. With this change, hydrogen demand is no longer an exogenous input, as hydrogen is one of several energy carriers and industrial feedstocks that the model can choose. Also, the availability of natural gas is modelled explicitly with available resources, LNG terminals, and pipeline capacities. An introduction to how EMPIRE is generally set up is given in \cref{app:EMPIRE}.

The two temporal scales in the multi-horizon framework are the long-term strategic periods, and the short-term operational hours. The strategic periods are each five years long, and EMPIRE can invest in new capacity for all assets at the start of each strategic period. The operational hours are linked to each strategic period, featuring hourly dispatch of the assets to meet the demand of each commodity, such as \eg, power. EMPIRE represents each of the meteorological seasons with one representative week of hourly operations each, as well as two days of peak power demand. This temporal resolution is to validate the investments made in the strategic period, and the operational costs for these representative weeks and peak days are scaled up to represent the operational cost for one representative year. EMPIRE features uncertainty in its operations, where each operational scenario consists of such a representative year. There are three such operational scenarios in this work, where the uncertain parameters include renewable power generation and electric power demand.

EMPIRE features 52 nodes to represent the European energy system. 30 nodes are for countries in Europe, in addition to 5 nodes for the five power price zones in Norway. There are also 14 offshore wind farm nodes, and one offshore energy hub node as in~\citet{Durakovic2023PoweringPrices}. The remaining two nodes are the Sleipner and Draupner offshore platforms, which are used to transport natural gas in the North Sea. The industries included in EMPIRE include the steel, cement, ammonia and oil refining industries, all of which have the potential for large-scale use of hydrogen in the future.

EMPIRE features a cap on annual CO$_2$ emissions, in line with the targets set by the~\citet{EuropeanCommission2018}. Whereas the European Commission separates the CO$_2$ emissions from the power and industry sectors, in EMPIRE, these separate caps are added into one shared cap for all sectors, giving the model the freedom to trade emissions across sectors if necessary.

Previously, natural gas was assumed to be abundant, and following the price as reported by the~\citet{EuropeanCommission2016EU2050}. This has changed in this work in order to reflect the lack of Russian natural gas in the energy system. Instead, the production and transmission of natural gas are now modelled explicitly in EMPIRE, where production can occur in Russia, North Africa or in the North Sea. No new pipeline capacity or liquid natural gas (LNG) import capacity can be built, where the existing pipeline capacity is taken from ENTSO-G as implemented by~\citet{Egging-Bratseth2021FreedomModel}, and the LNG capacity of each country is as reported by~\citet{GasInfrastructureEurope2022GIEDatabase}. All reserves estimates are taken from~\citet{bp2021Bp2021}, except for the Norwegian reserves, which are allocated to the three south-western price zones based on geographic location as reported by~\citet{NorwegianPetroleum2023Fields}. Furthermore, in the cases where Russian gas is included, it is assumed that there is an unlimited supply from Russia, and the only limiting factor is the pipeline capacity. Similarly, LNG supply is also assumed to be inexhaustible, and is only limited by the import capacity. The production capacity of Norway is split into the three power price zones, where the production capacity of the price zone is the capacity of Kårstø~\cite{Equinor2023Landanlegg} in NO2, of Nyhamna~\cite{Gassco2023NyhamnaPlant} in NO3, and of Kollsnes~\cite{Equinor2023Landanlegg} in NO5. The natural gas production capacity of the UK was taken from the~\citet{EnergyInformationAdministration2022CountryKingdom}. The natural gas coming from North Africa is assumed to be constrained by the pipeline capacities into Spain and Italy, and so these are the limits for this source. To represent the flexibility in the North Sea pipeline network, the two hub platforms Sleipner and Draupner are also represented, thereby representing the North Sea gas pipeline network similarly to~\citet{Kazda2020OptimalUncertainty}. These are initially powered by on-site gas turbines, and have the option of electrification from mainland Norway. Some countries also have long-term natural gas storage, with the total capacity for this taken from the~\citet{EuropeanCommission2022QuestionsStorage}.

Cost of producing natural gas is assumed to be the same in the North Sea, Russia and North Africa, and every country is assumed to pay the same price for LNG. These prices are uncertain, and so two cases a constructed, where the natural gas is more costly in one case. In the affordable case, natural gas production is assumed to cost 10~EUR/MWh, and in the costly case, this cost is doubled to 20~EUR/MWh. The LNG prices are summarized in \cref{tab:lng_price}.

\begin{table}[tb!]
    \centering
    \begin{tabular}{lcc} 
    \hline
         \textbf{Year} & \textbf{Affordable LNG} & \textbf{Costly LNG} \\
         2020 & 20.86 & 50.98\\
         2025 & 22.57 & 55.15\\
         2030 & 24.55 & 59.98\\
         2035 & 26.22 & 64.06\\
         2040 & 27.10 & 66.22\\
         2045 & 27.66 & 67.57\\
         2050 & 28.08 & 68.62\\
         2055 & 28.08 & 68.62\\
    \hline
    \end{tabular}
    \caption{Price for LNG in affordable and costly case}
    \label{tab:lng_price}
\end{table}

The CCS chain is modelled such that CO$_2$ can be captured from certain power generators fuelled by coal or natural gas, from hydrogen production with natural gas reforming and from certain industry plants, when applied in the steel and cement sectors. CO$_2$ can be transported internationally using pipelines, and can only be permanently sequestered in the North Sea. \cref{tab:co2_sequestration capacity} shows which nodes can sequester CO$_2$ in this work, and the corresponding maximum capacity for sequestration.

\begin{table}[tb!]
    \centering
    \begin{tabular}{lrc}
    \hline
         \multirow{2}{*}{\textbf{Node}} & \textbf{CO$_2$ sequestration } & \multirow{2}{*}{\textbf{Reference}}\\
         & \textbf{capacity [Gt]} & \\ 
         NO2 & 29.5 &  \cite{Halland2022CO2Sea}\\
         NO3 & 30.7 &  \cite{Halland2022CO2Sea}\\
         NO5 & 0.2 & \cite{Halland2022CO2Seab}\\
         Denmark & 0.3 & \cite{Turgut2021AnEurope} \\
         The Netherlands & 4.0 & \cite{Turgut2021AnEurope}\\
         Great Britain & 78.0 & \cite{Turgut2021AnEurope}\\ 
     \hline
    \end{tabular}
    \caption{Maximum capacity for offshore CO$_2$ sequestration in the North Sea.}
    \label{tab:co2_sequestration capacity}
\end{table}



The industry is represented by the steel, cement, ammonia and oil refining industries. The yearly output of steel is taken from the~\citet{EuropeanParliamentaryResearchService2021Carbon-freeInfrastructure}, where the future growth is assumed to follow the growth trajectory as reported by the~\citet{InternationalEnergyAgency2020IronSeries}. The initial capacity of each country is taken from~\citet{EUROFER2019MapSites}. It is assumed that the total scrap use cannot exceed 45\% of the total annual crude steel demand, which is roughly the average share of electric arc furnace production in Europe from 2012 to 2021~\cite{EUROFER2022EuropeanFigures}.

Ammonia demand is taken from \citet{Egenhofer2014AAmmonia}. For the initial capacity, it is assumed that this demand is met as if all of it were produced by ammonia plants with local steam methane reforming (SMR) without CCS, and that the capacities of these initial plants are as if they meet the yearly demand by producing at maximum capacity all year. The alternative way to produce ammonia in this model is to have an ammonia plant that receives hydrogen from the hydrogen market rather than producing it locally. 

Cement is another sector that can potentially benefit from hydrogen and CCS, especially the latter as the decomposition of limestone to calcium oxide in clinker production emits roughly 0.78 tons of CO$_2$ per ton of clinker produced. These emissions also occur even if the fuel in the kiln is completely emissions free. In this model, the yearly demand for cement is taken from the~\citet{USGeologicalSurvey20212016Eurasia}, where the clinker to cement ratio is assumed to be improved as described by the~\citet{InternationalEnergyAgency2018TechnologyIndustry}. The present capacity is assumed to be such that the yearly demand is met by the initial capacity is run at maximum capacity all year long.

Refineries consume significant amounts of hydrogen and are included here as an industrial sector. \citet{McKinseyInsights2022EuropeanRefineries} gives the refinery production capacities of each European country, which is used to meet the demand for refined oil. This demand is falling as Europe is decarbonizing, and the yearly demand for refined oil is decreased based on the decrease of refinery runs in as reported by the~\citet{InternationalEnergyAgency2021World2021}.


The transport sector is modelled in a simplified way such that the annual energy demand for each energy carrier, as reported by the~\citet{EuropeanCommission2020EU2020}, is met. The transport sector is thus an exogenous demand, and the model makes no decisions about the technologies that are used.

The full code and all data is available as open access on the public project Github page~\cite{Durakovic2023EmpirePublic}.

\section{Results and analysis} \label{sec:results}

This section includes the results and analysis of these. Four cases are considered, featuring the different permutations of affordable and costly natural gas, and with and without Russian natural gas. \cref{sec:power_heat} focuses on the temporal development of the power and domestic heat sectors, \cref{sec:hydrogen} analyzes how the development of hydrogen production changes between the different cases, \cref{sec:industry} shows the changes in industrial production for the cement and steel industries, and finally, \cref{sec:co2} shows the utilization of CCS.

\subsection{Power \& domestic heat sectors}
\label{sec:power_heat}

\begin{figure}[ht!]
    \centering
    \begin{subfigure}[htb]{0.49\textwidth}
        \centering
        \includegraphics[width=\textwidth]{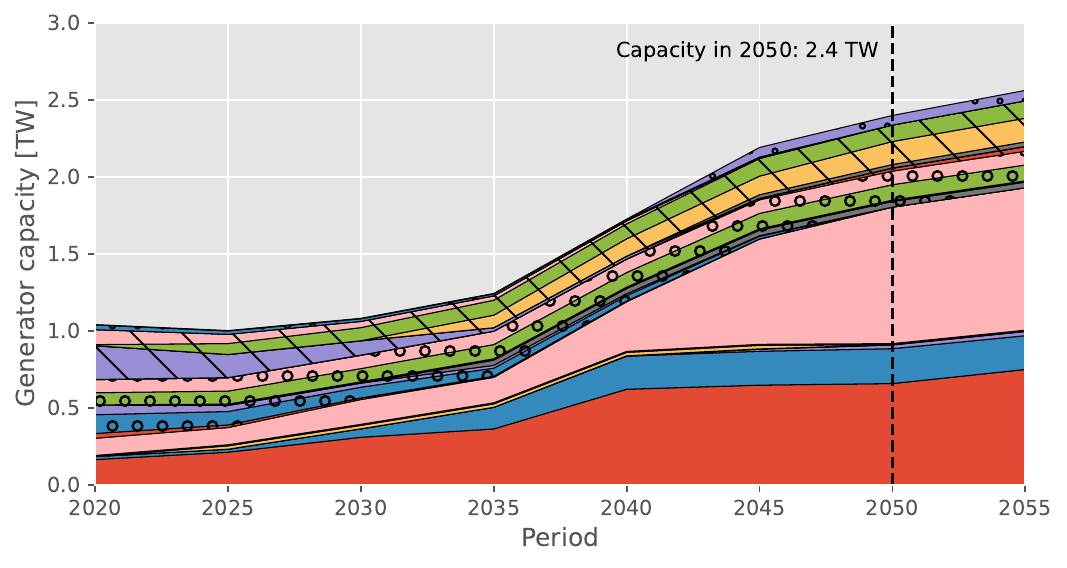}     
        \caption{Affordable, with Russian gas}
        \label{fig:power_ru_cheap}
    \end{subfigure}
    \begin{subfigure}[htb]{0.49\textwidth}
        \centering
        \includegraphics[width=\textwidth]{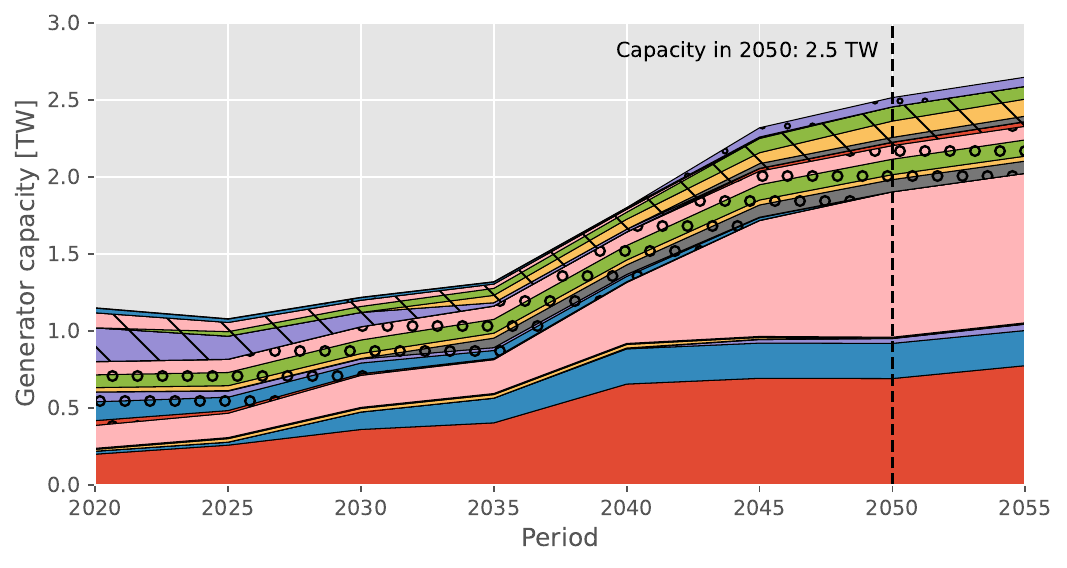}     
        \caption{Costly, with Russian gas}
        \label{fig:power_ru_exp}
    \end{subfigure}
    \\
    \begin{subfigure}[htb]{0.49\textwidth}
        \centering
        \includegraphics[width=\textwidth]{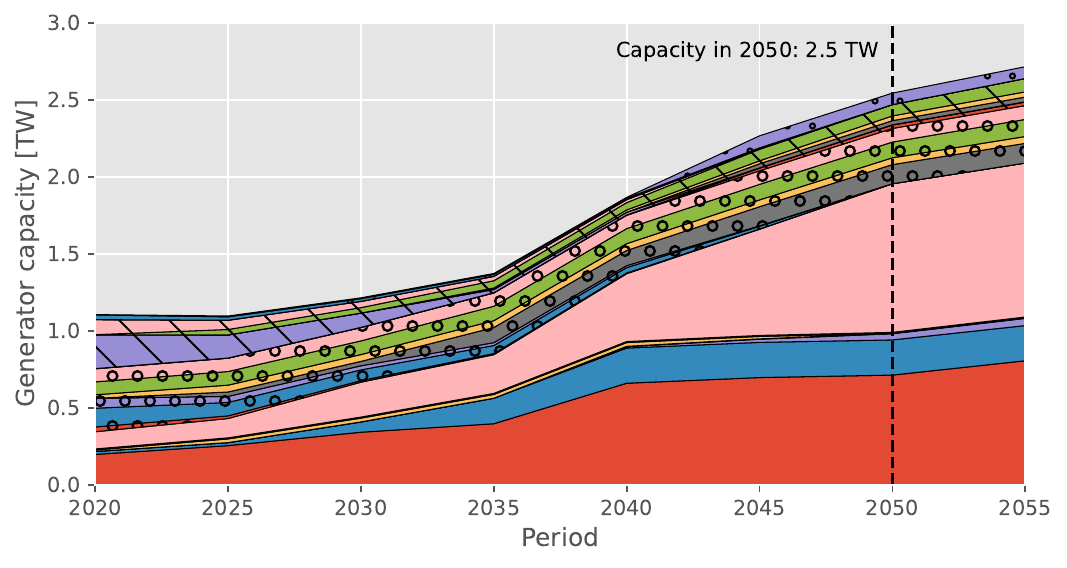}     
        \caption{Affordable, without Russian gas}
        \label{fig:power_cheap}
    \end{subfigure}
    \begin{subfigure}[htb]{0.49\textwidth}
        \centering
        \includegraphics[width=\textwidth]{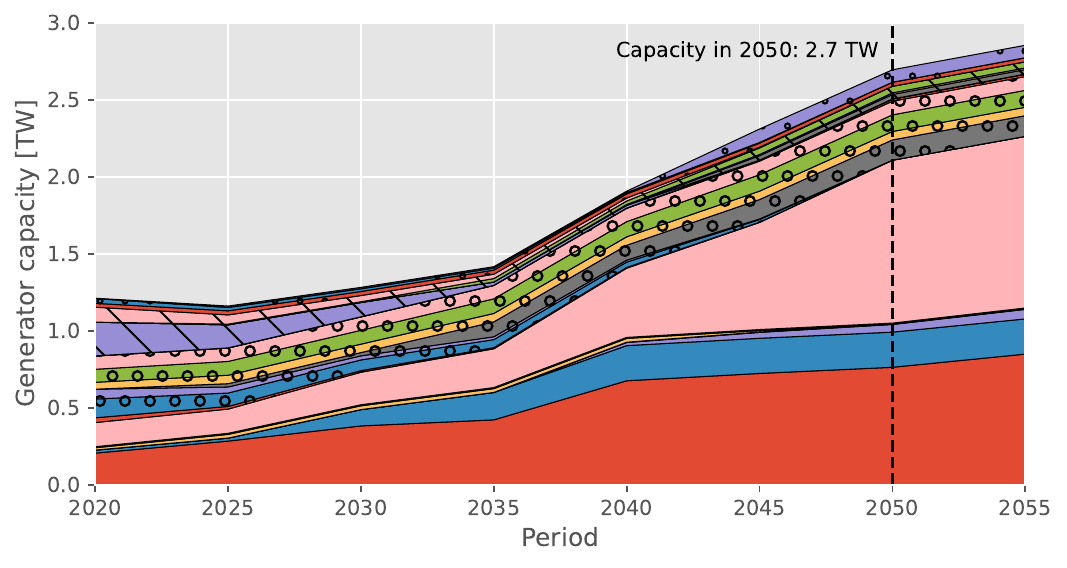}     
        \caption{Costly, without Russian gas}
        \label{fig:power_exp}
    \end{subfigure}
    \\
    \begin{subfigure}[htb]{0.9\textwidth}
        \centering
        \includegraphics[width=\textwidth]{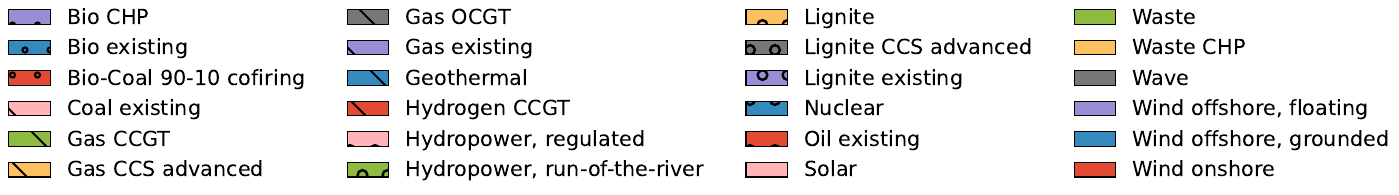}
    \end{subfigure}
    \caption{Development of European power sector.}
    \label{fig:power}
\end{figure}

The European power demand is predicted to increase considerably in conjunction with tightening CO$_2$ emission caps. \cref{fig:power} shows the development of the European power generation capacity, subject to these two developments.

The four cases shown in \cref{fig:power} share some similarities. The first is that there is a large growth in power generation capacity in Europe, by at least 130\% between 2020 and 2050. The second important observation is that this growth is mainly underpinned by the renewable generators of solar and wind. Furthermore, both onshore and offshore wind play large roles in the power system in 2050, where grounded offshore wind accounts for most of the offshore wind capacity, but floating offshore wind still has between 24.0 and 49.7~GW of capacity, depending on the case. Renewable power generators are thus at the core of the European power sector, with other dispatchable generators supplementing the renewables when there is insufficient renewable power generation to meet all demand. All four cases also feature hydrogen-fuelled power generators, but these only play a minor role, where the capacities total capacity for hydrogen-fuelled generators in 2050 is between 13.0 and 22.4~GW.

There are also some important differences between the cases in \cref{fig:power}. One trend that can be observed is how the total power generation capacity grows as access to natural gas is restricted, either through higher natural gas costs, or by removing Russian gas. Comparing the most relaxed case in \cref{fig:power_ru_cheap} with the most restrictive case in \cref{fig:power_exp}, it can be seen that the total power generation capacity in 2050 grows from 2.4~TW to 2.7~TW, or by about 12.5\%. It can also be observed how the total installed generation capacities of the renewable generators grow considerably as access to natural gas is restricted, with onshore solar and wind having the largest increase.

Another important difference between the four cases is the role of natural gas in the power sector. In \cref{fig:power_ru_cheap,fig:power_ru_exp} natural gas power generators, both with and without CCS, account for a significant share of the power generation capacity, whereas in \cref{fig:power_cheap,fig:power_exp} these capacities are strongly diminished. The power system requires dispatchable power that gas-powered generators previously offered, and in \cref{fig:power_cheap,fig:power_exp}, this role is filled by coal-fired power plants with CCS. Furthermore, as previously discussed, renewables account for a larger share of the power generation capacity.



\begin{figure}[ht!]
    \centering
    \begin{subfigure}[htb]{0.49\textwidth}
        \centering
        \includegraphics[width=\textwidth]{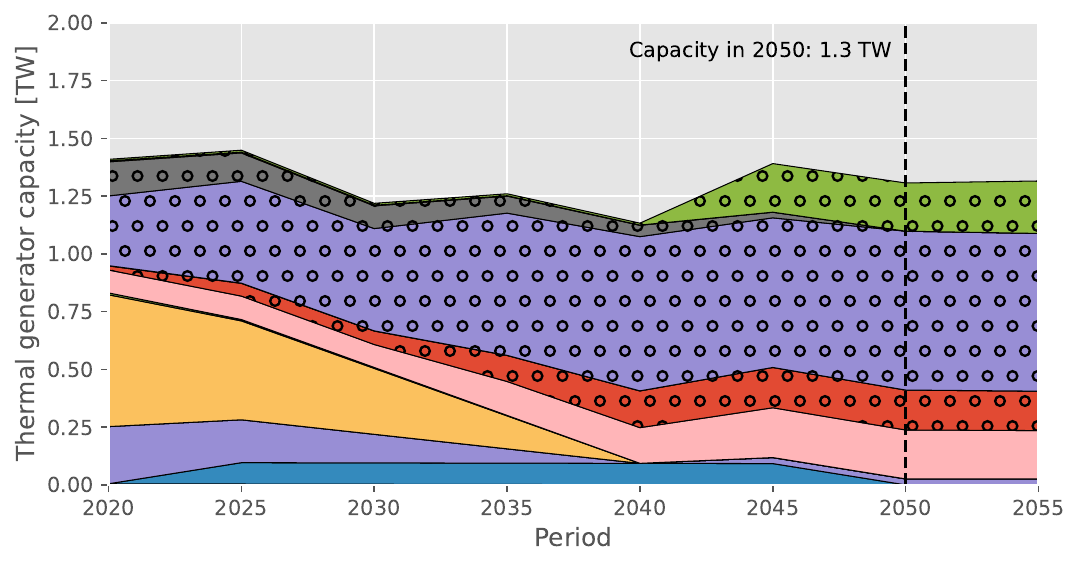}     
        \caption{Affordable, with Russian gas}
        \label{fig:heat_ru_cheap}
    \end{subfigure}
    \begin{subfigure}[htb]{0.49\textwidth}
        \centering
        \includegraphics[width=\textwidth]{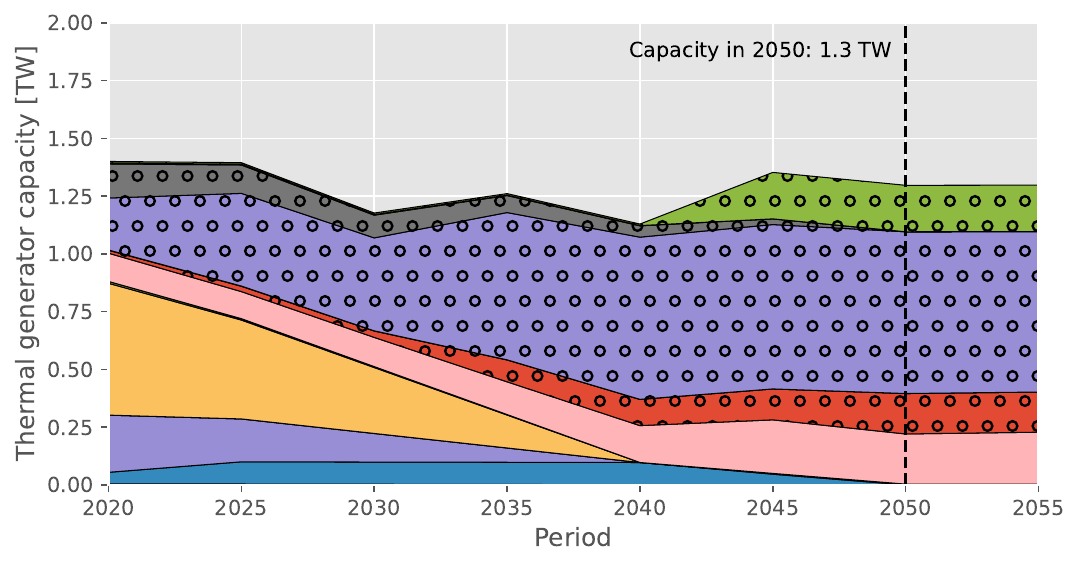}     
        \caption{Costly, with Russian gas}
        \label{fig:heat_ru_exp}
    \end{subfigure}
    \\
    \begin{subfigure}[htb]{0.49\textwidth}
        \centering
        \includegraphics[width=\textwidth]{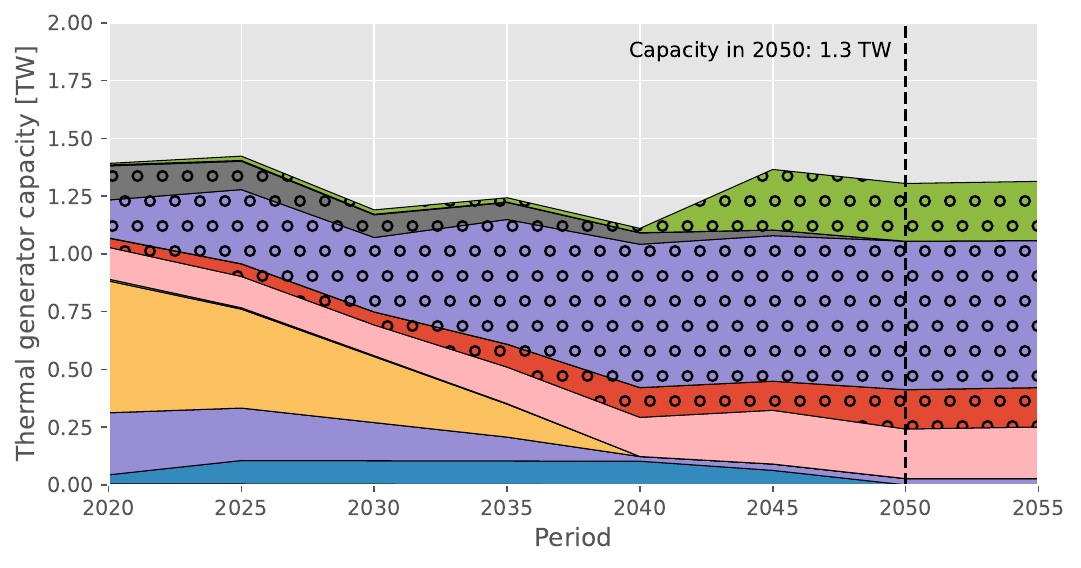}     
        \caption{Affordable, without Russian gas}
        \label{fig:heat_cheap}
    \end{subfigure}
    \begin{subfigure}[htb]{0.49\textwidth}
        \centering
        \includegraphics[width=\textwidth]{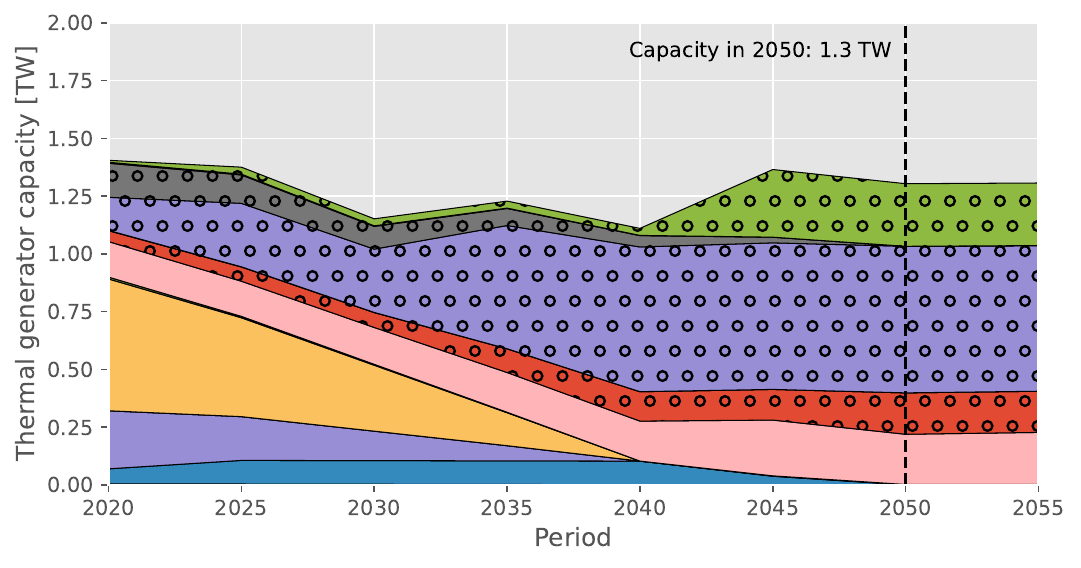}     
        \caption{Costly, without Russian gas}
        \label{fig:heat_exp}
    \end{subfigure}
    \\
    \begin{subfigure}[htb]{0.9\textwidth}
        \centering
        \includegraphics[width=\textwidth]{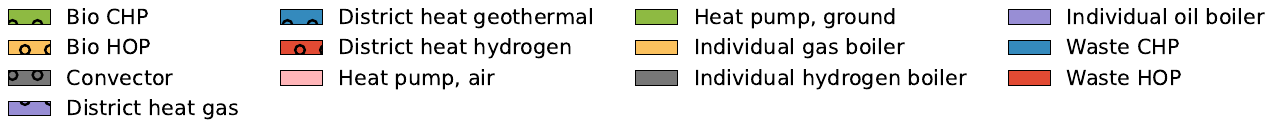}
    \end{subfigure}
    \caption{Development of European domestic heat sector.}
    \label{fig:heat}
\end{figure}

\cref{fig:heat} shows the development of the European domestic heat sector in the four cases. Overall, the development is very similar in all cases. It can be observed how the domestic heat sector tends towards larger centralized combined heat and power (CHP) and district heat systems. The decentralized gas-based systems are simultaneously phased out. There is also a pivot towards individual air-source heat pumps, as opposed to boilers for individual households. Note that the capacity shown for heat pumps in \cref{fig:heat} is the electric capacity of the heat pump, as the coefficient of performance is stochastic, depending on the outside temperature in each country. The coefficients of performance are between 1.83 and 3.33. The heat output of the heat pump systems is thus higher than suggested by \cref{fig:heat}.

Inspecting the differences between the cases, it can be seen how when access to Russian gas is removed, then there is a larger reliance on bioenergy-based CHP plants, where the capacity in 2050 increases by at least 18\% compared to the respective case with Russian gas. The expanded use of heat pumps is in line with the results presented by~\citet{Pedersen2022}, but the use of hydrogen in the heat sector is not. In their results, hydrogen is not used in the heating sector at all, while electricity-based heat, from both heat pumps and resistive heating, is used extensively. One reason for this difference may be that their model is deterministic, thereby potentially overestimating the availability of renewable electricity. Taking the uncertainty of renewable generation into account has been shown to favor dispatchable generators~\cite{Seljom2015b}.

In short, \cref{fig:power,fig:heat} show how energy production for both power and heat rely more on energy sources within the EU, in terms of renewable energy generation, bioenergy, and to some extent, coal. This comes at the expense of gas use, which was previously in large part sourced from Russia.

\subsection{Hydrogen production}
\label{sec:hydrogen} 

\begin{figure}[ht!]
    \centering
    \begin{subfigure}[htb]{0.49\textwidth}
        \centering
        \includegraphics[width=\textwidth]{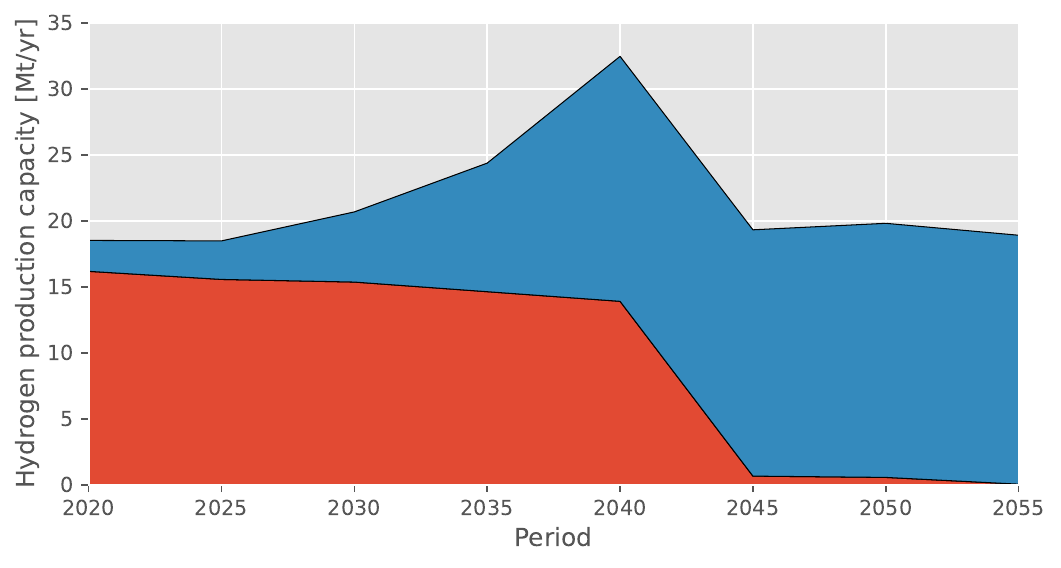}     
        \caption{Affordable, with Russian gas}
        \label{fig:h2_ru_cheap}
    \end{subfigure}
    \begin{subfigure}[htb]{0.49\textwidth}
        \centering
        \includegraphics[width=\textwidth]{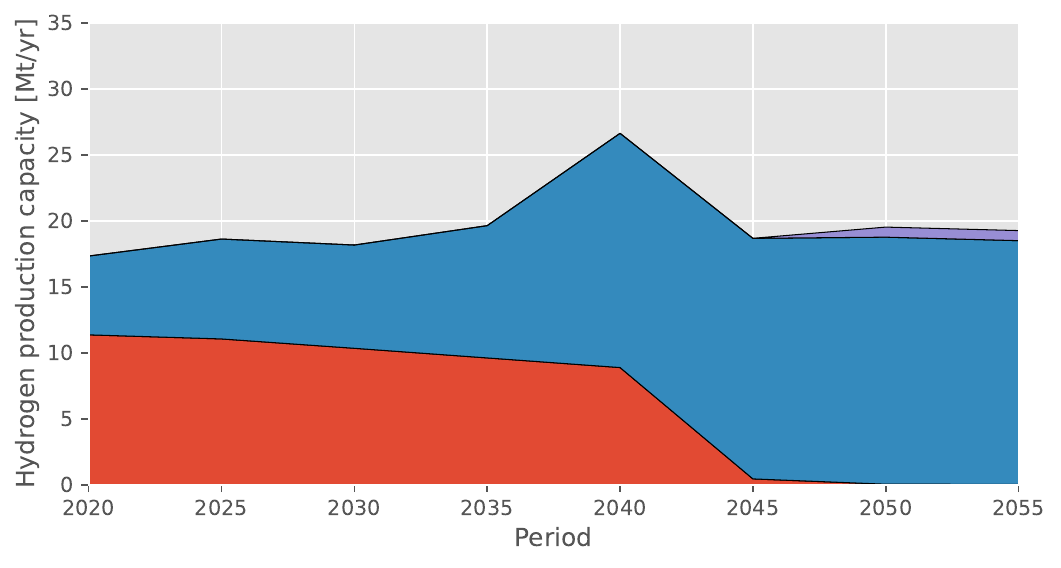}     
        \caption{Costly, with Russian gas}
        \label{fig:h2_ru_exp}
    \end{subfigure}
    \\
    \begin{subfigure}[htb]{0.49\textwidth}
        \centering
        \includegraphics[width=\textwidth]{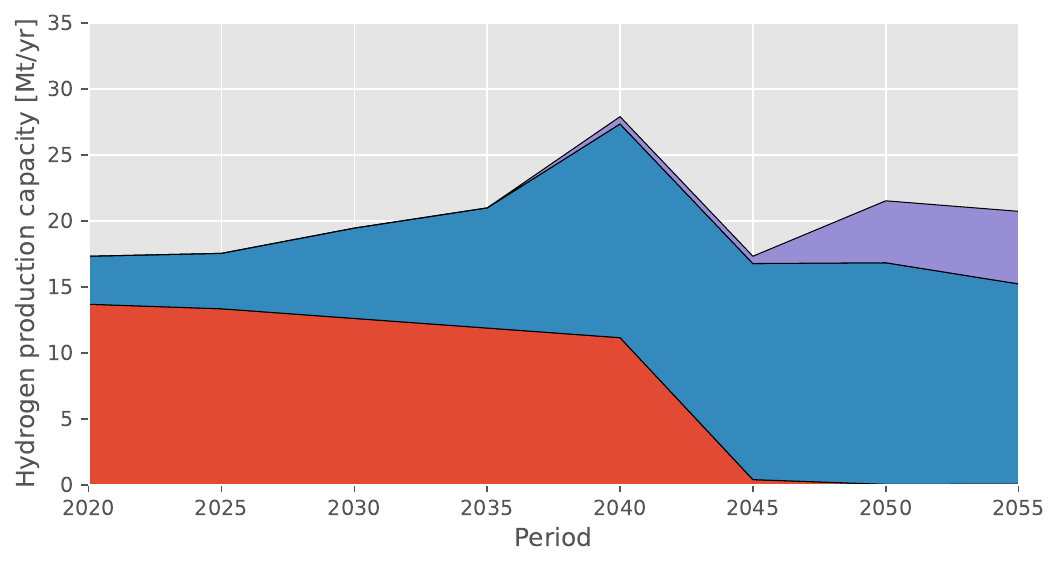}     
        \caption{Affordable, without Russian gas}
        \label{fig:h2_cheap}
    \end{subfigure}
    \begin{subfigure}[htb]{0.49\textwidth}
        \centering
        \includegraphics[width=\textwidth]{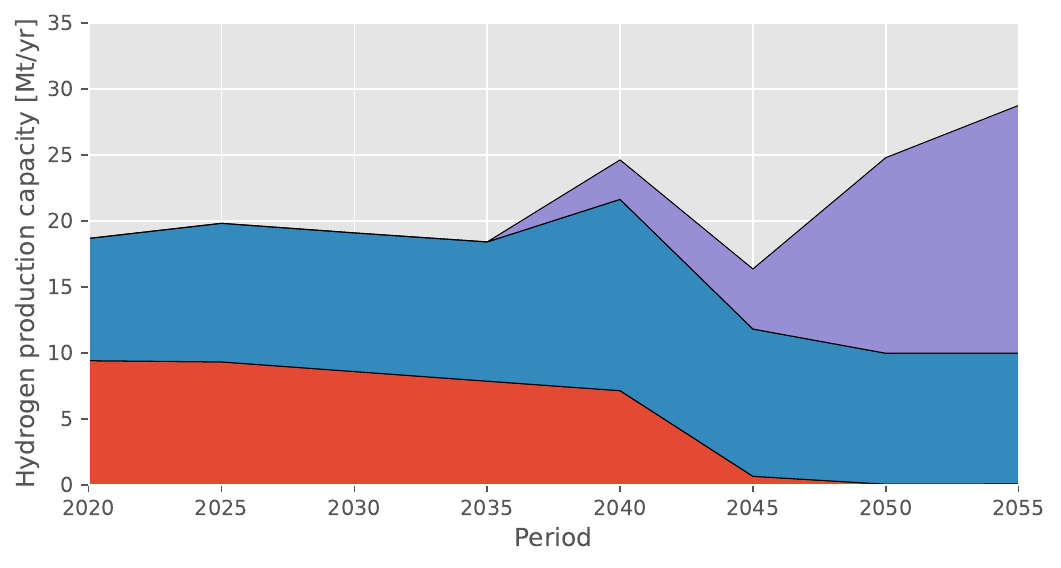}     
        \caption{Costly, without Russian gas}
        \label{fig:h2_exp}
    \end{subfigure}
    \\
    \begin{subfigure}[htb]{0.9\textwidth}
        \centering
        \includegraphics[width=\textwidth]{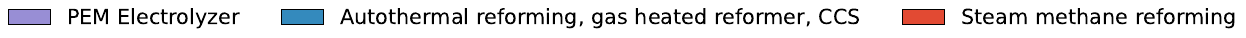}
    \end{subfigure}
    \caption{Development of hydrogen production capacity in Europe.}
    \label{fig:h2}
\end{figure}

Hydrogen is an important energy carrier in a decarbonized energy system, where it can be used to decarbonize power and heat supply, as well as energy and feedstock supply in industry. Hydrogen is also used in the exogenous energy demand in transport, which has to be met in this model. \cref{fig:h2} shows the development of hydrogen production, as it is decarbonized along with the rest of the energy system, including the locally produced hydrogen for ammonia production, which is included in the steam methane reforming group. 

\cref{fig:h2_ru_cheap,fig:h2_ru_exp} show that when Russian gas is available, the most cost-effective way to produce hydrogen is through natural gas reforming. In the beginning, this hydrogen production is mainly based on SMR without CCS, much like hydrogen production today, but this way of producing hydrogen is substituted by autothermal reforming in the long term, using gas heated reformers for improved efficiency and CCS for reduced CO$_2$ emissions. 

Interestingly, there is no substantial electrolyzer capacity in either the affordable or costly case when Russian gas is available. This is because there is an abundance of affordable pipeline gas, and the included technologies are able to produce hydrogen with a very high CO$_2$ capture rate, allowing for the production of hydrogen with very low greenhouse gas emissions. At the same time, it is assumed that the delivered natural gas is not associated with any methane leak, whereas in reality, certain countries have considerable methane emissions associated with natural gas production, including for example Russia and Algeria~\cite{InternationalEnergyAgency2023GlobalOperations}. Accounting for the greenhouse effect from these methane leaks can have a significant impact on the climate footprint of blue hydrogen~\cite{Howarth2021HowHydrogen}, which can significantly influence these results by facilitating an increased production of green hydrogen. In considering the greenhouse effect from methane leaks, it is important to differentiate on where the hydrogen comes from~\cite{Romano2022CommentHydrogen}, advantaging Norwegian blue hydrogen. These results are aligned with the 2022 report by \citet{Hydrogen4EU2022Hydrogen4EU:Edition}, where upstream methane leak was considered in the development of a hydrogen supply chain. In this report, the distribution between blue and green in their \textit{Technology Diversification} case was similar to what is seen in \cref{fig:h2_exp}, emphasizing the potential of blue hydrogen production.

Removing access to Russian gas, as in~\cref{fig:h2_cheap,fig:h2_exp}, leads to some important differences. While the development of hydrogen production capacity looks similar in the short timeframe, it can also be observed how green hydrogen plays a much more important role in these cases, especially in~\cref{fig:h2_exp} where natural gas is costly. In these cases, there is substantially less pipeline gas available in the market, and much of the natural gas demand is met through LNG imports. In the case shown in~\cref{fig:h2_cheap}, the LNG is affordable enough that it is economical to produce blue hydrogen from LNG imports. However, in the costly gas case, this occurs much more rarely, and pipeline gas is the main source of natural gas for hydrogen production. Since there is much less pipeline gas available in the case shown in~\cref{fig:h2_exp}, it becomes much more attractive to produce hydrogen through electrolysis. By 2050, green hydrogen accounts for almost 60\% of the total hydrogen production capacity in~\cref{fig:h2_exp}, as the green hydrogen production capacity increases in conjunction with the large increase of renewable power capacity seen in~\cref{fig:power_exp}.

In the REPowerEU plan~\cite{EuropeanCommission2022REPowerEUPlan}, the European Commission set a goal of 20~Mt of annual renewable hydrogen production, with 10~Mt being produced inside the EU, and the remaining 10~Mt being imported from nearby regions. None of the results shown in~\cref{fig:h2} reach this goal. Instead, by 2030, all of the hydrogen production capacity is in natural gas reforming, and with the majority being SMR without CCS. Most of this capacity comes from local hydrogen production for ammonia. Considering the development of the power sector as shown in~\cref{fig:power}, it is evident that by 2030 there is not enough renewable power to support 20~Mt of renewable hydrogen production. In order to achieve these goals, it is therefore necessary to build up a much larger capacity of renewable power generation by 2030.
At the same time, the results suggest that this may not be necessary; it is possible to reach the carbon neutrality goals without needing 20~Mt of renewable hydrogen in 2030, and also without relying on Russian gas.

\subsection{Industry}
\label{sec:industry}

This work includes the steel, cement and ammonia industries in order to cover their hydrogen demand, and see to what degree they use CCS, when possible. 

\begin{figure}[ht!]
    \centering
    \begin{subfigure}[htb]{0.49\textwidth}
        \centering
        \includegraphics[width=\textwidth]{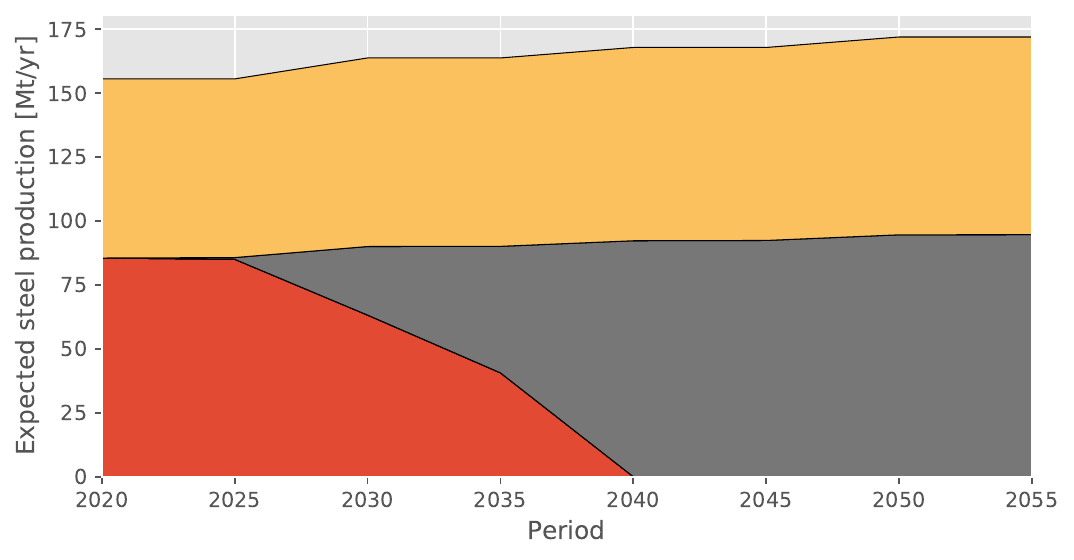}     
        \caption{Affordable, with Russian gas}
        \label{fig:steel_ru_cheap}
    \end{subfigure}
    \begin{subfigure}[htb]{0.49\textwidth}
        \centering
        \includegraphics[width=\textwidth]{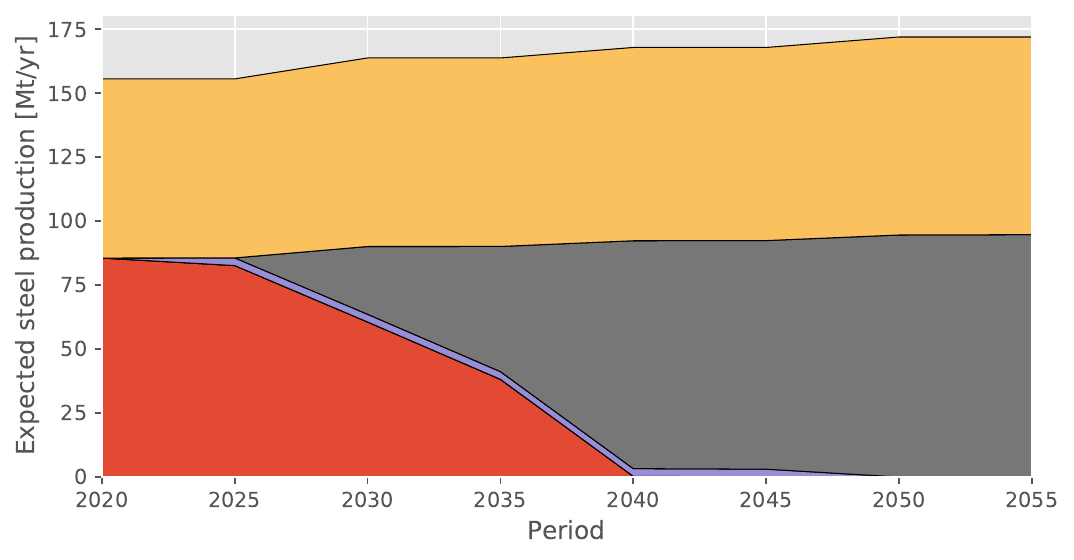}
        \caption{Costly, with Russian gas}
        \label{fig:steel_ru_exp}
    \end{subfigure}
    \\
    \begin{subfigure}[htb]{0.49\textwidth}
        \centering
        \includegraphics[width=\textwidth]{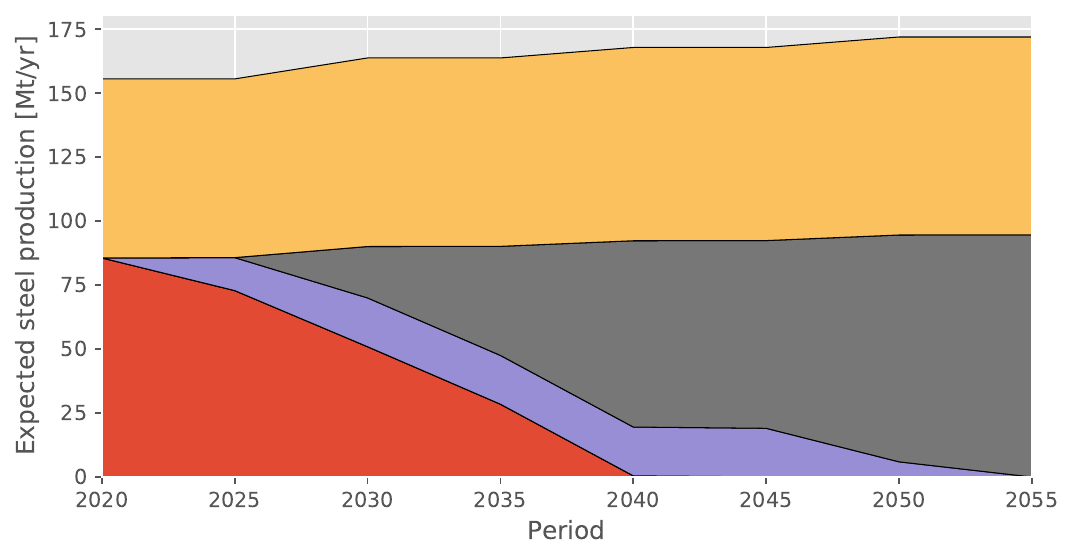}     
        \caption{Affordable, without Russian gas}
        \label{fig:steel_cheap}
    \end{subfigure}
    \begin{subfigure}[htb]{0.49\textwidth}
        \centering
        \includegraphics[width=\textwidth]{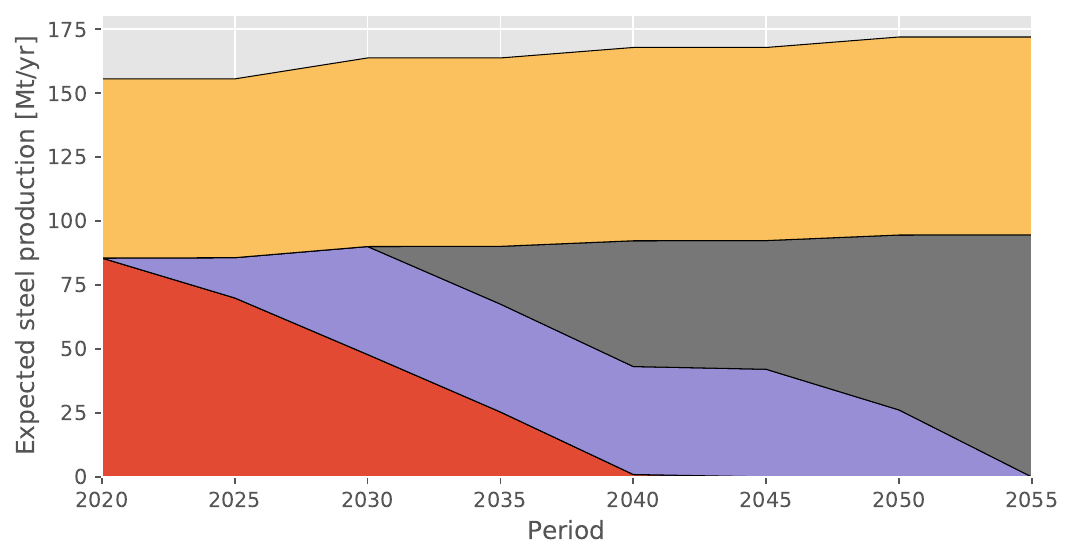}     
        \caption{Costly, without Russian gas}
        \label{fig:steel_exp}
    \end{subfigure}
    \\
    \begin{subfigure}[htb]{0.6\textwidth}
        \centering
        \includegraphics[width=\textwidth]{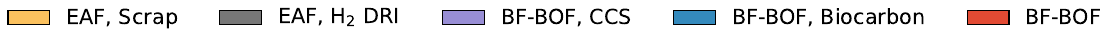}
    \end{subfigure}
    \caption{Evolution of European steel production.}
    \label{fig:steel}
\end{figure}

\cref{fig:steel} shows the production of European steel, and the share of the total production that each steel plant accounts for. Common for all four cases is how the use of scrap is maximized, as this way of producing steel is emissions-free. \cref{fig:steel} also shows how eventually, regardless of the case, all European steel is made in electric arc furnace (EAF) plants, that either use scrap or iron reduced using hydrogen as a feedstock. Biocarbon is also not used in any of the cases, instead favoring CCS and hydrogen as decarbonization pathways.

The difference in the four cases mainly occurs as Russian gas is removed in~\cref{fig:steel_cheap,fig:steel_exp}. While also in these cases, steel production ultimately relies completely on hydrogen and scrap, the transition to this final state is different than the cases seen in~\cref{fig:steel_ru_cheap,fig:steel_ru_exp}. Whereas the cases with Russian gas transition directly from the conventional blast furnace, basic oxygen furnace (BF-BOF) technology to hydrogen direct reduced iron with EAF, the cases without Russian gas go through an intermediate step with steel plants using the BF-BOF technology, but with CCS. This comes as a result of there being less affordable hydrogen available in the energy system when the Russian gas is removed; it becomes more effective to decarbonize through CCS while the hydrogen market matures, even though the CO$_2$ capture rate in the steel sector is relatively low at 60\%. In this way, the steel industry avoids having to use relatively scarce natural gas (through the consumption of blue hydrogen), and can instead continue using the more abundant coal.

\begin{figure}[ht!]
    \centering
    \begin{subfigure}[htb]{0.49\textwidth}
        \centering
        \includegraphics[width=\textwidth]{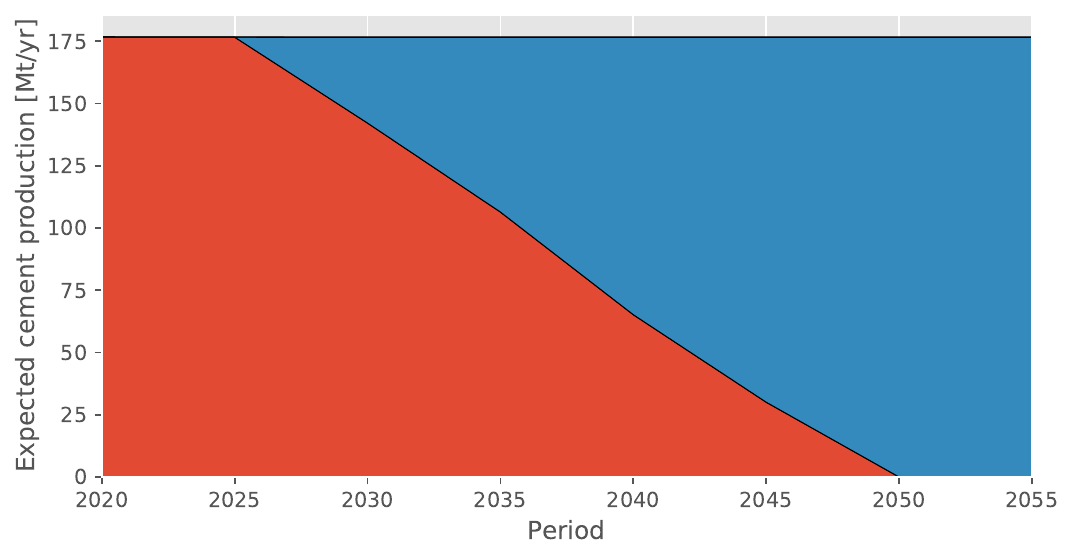}     
        \caption{Affordable, with Russian gas}
        \label{fig:cement_ru_cheap}
    \end{subfigure}
    \begin{subfigure}[htb]{0.49\textwidth}
        \centering
        \includegraphics[width=\textwidth]{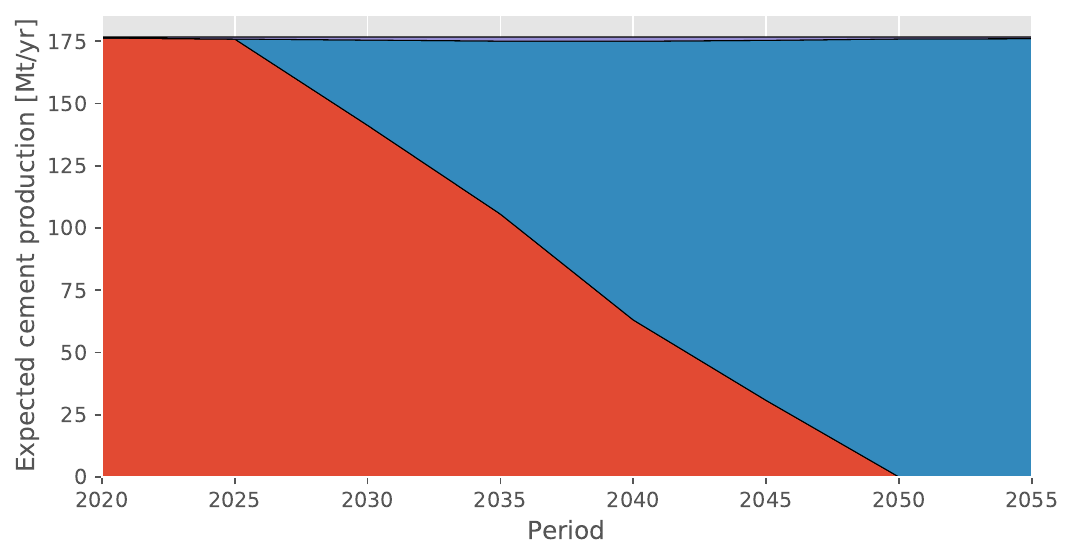}
        \caption{Costly, with Russian gas}
        \label{fig:cement_ru_exp}
    \end{subfigure}
    \\
    \begin{subfigure}[htb]{0.49\textwidth}
        \centering
        \includegraphics[width=\textwidth]{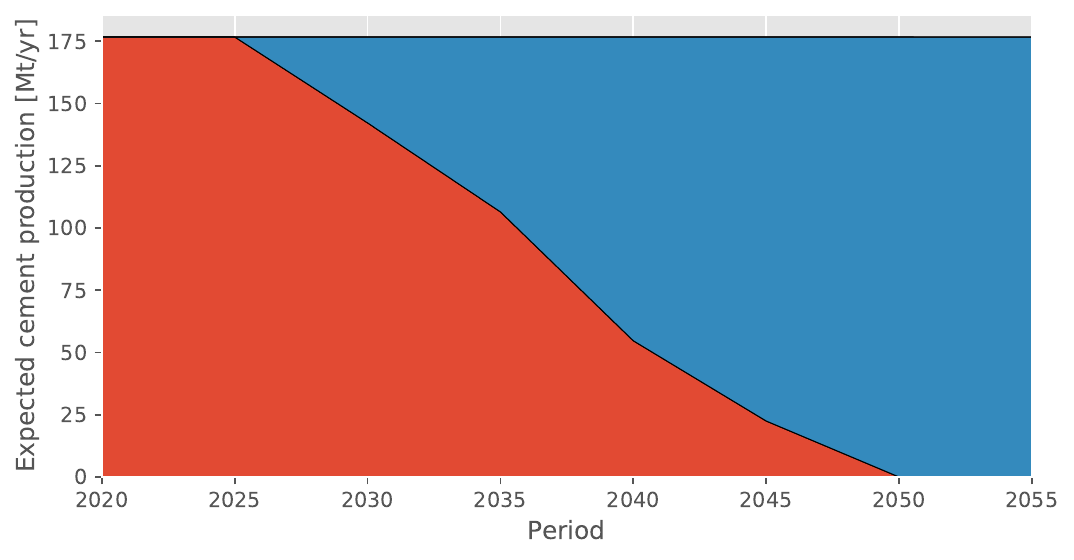}     
        \caption{Affordable, without Russian gas}
        \label{fig:cement_cheap}
    \end{subfigure}
    \begin{subfigure}[htb]{0.49\textwidth}
        \centering
        \includegraphics[width=\textwidth]{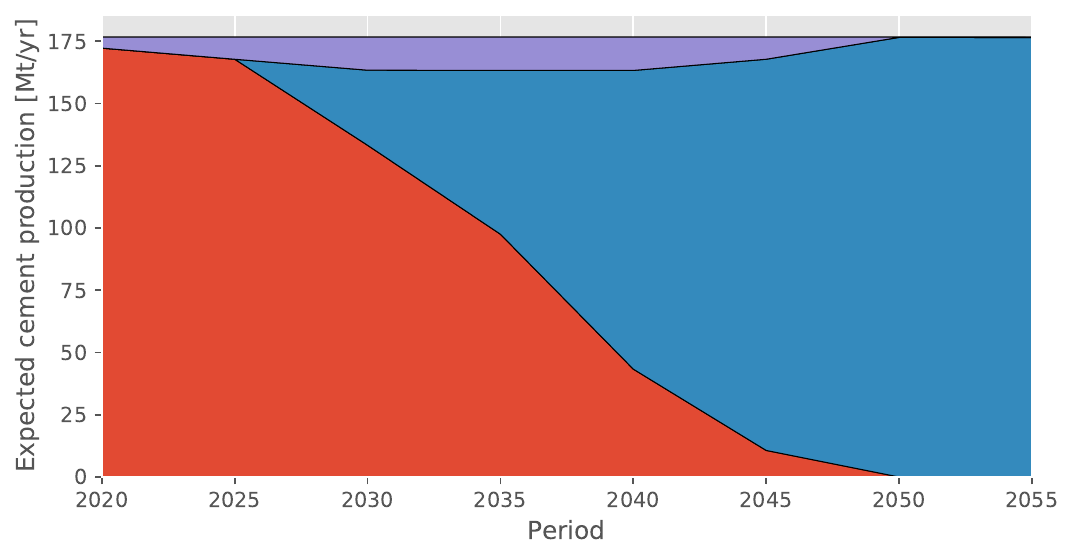}     
        \caption{Costly, without Russian gas}
        \label{fig:cement_exp}
    \end{subfigure}
    \\
    \begin{subfigure}[htb]{0.6\textwidth}
        \centering
        \includegraphics[width=\textwidth]{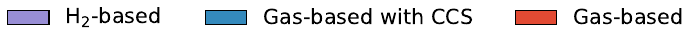}
    \end{subfigure}
    \caption{Evolution of European cement clinker production.}
    \label{fig:cement}
\end{figure}

In this work, the cement industry can be decarbonized by building cement plants where the clinker is produced using gas while capturing the CO$_2$ emissions, or partially decarbonized by switching the fuel used in clinker production to hydrogen. \cref{fig:cement} shows how these three options decarbonize the cement industry.

Comparing the four cases, it can be observed how their developments in the cement sector appear almost identical. Starting from 2030, the cement sector is gradually decarbonized by introducing CCS to cement plants, and by 2050, all cement plants feature CCS in all four cases. This result is in line with what is presented by the~\citet{InternationalEnergyAgency2018TechnologyIndustry}, where CCS appears as a priority for the decarbonization of cement. 

In \cref{fig:cement_exp}, it can also be observed how a small share of clinker production experiences a fuel switch from natural gas to hydrogen before 2050. This result is counter-intuitive, as hydrogen production is largely based on natural gas, as seen in \cref{fig:h2_exp}, and the production of this hydrogen includes an efficiency loss, thereby ostensibly introducing inefficiencies in the energy system. The reason for this fuel switch is a modelling anomaly. The hydrogen-based cement plants in the results are constructed in south-eastern Europe, a region that has previously been supplied by Russian gas. The availability of this gas is removed in this case. Furthermore, a modelling assumption is that the model cannot build new natural gas pipelines, whereas it can build new hydrogen pipelines. As Russian gas is removed and LNG is (prohibitively) expensive, the existing natural gas pipeline infrastructure is not sufficient to sustain all the natural gas demand here. The model is thus forced to build hydrogen pipelines instead in order to meet the demand. This will in reality likely not develop as shown in \cref{fig:cement_exp}, as the infrastructure may be operated in a more efficient way that is not modelled, or if necessary, the gas infrastructure may be reinforced to suit the needs of the energy system.



\subsection{Sequestration of \texorpdfstring{CO$_2$}{CO2}}
\label{sec:co2}

\begin{figure}[ht!]
    \centering
    \begin{subfigure}[htb]{0.49\textwidth}
        \centering
        \includegraphics[width=\textwidth]{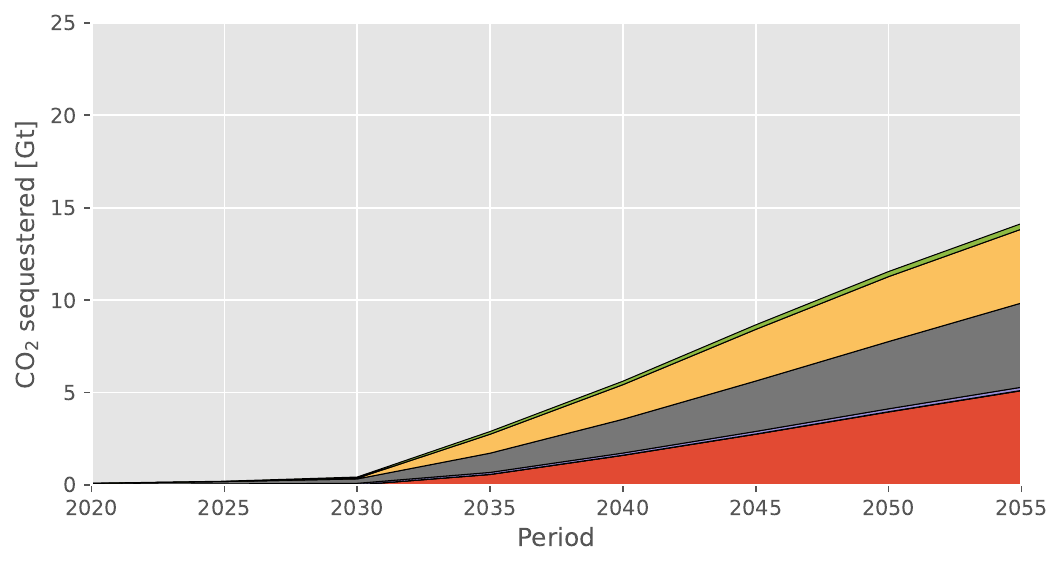}     
        \caption{Affordable, with Russian gas}
        \label{fig:co2_seq_ru_cheap}
    \end{subfigure}
    \begin{subfigure}[htb]{0.49\textwidth}
        \centering
        \includegraphics[width=\textwidth]{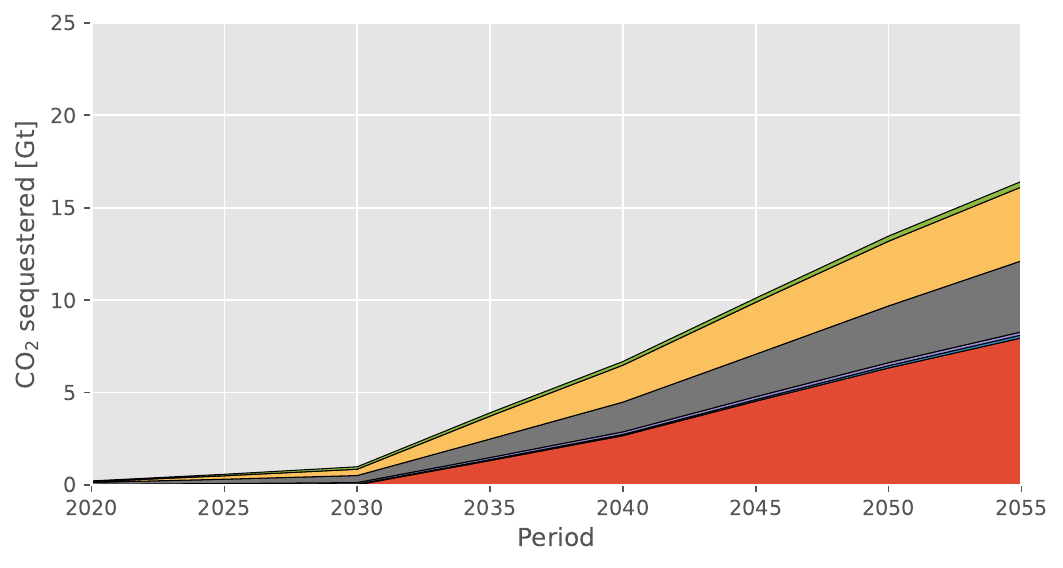}
        \caption{Costly, with Russian gas}
        \label{fig:co2_seq_ru_exp}
    \end{subfigure}
    \\
    \begin{subfigure}[htb]{0.49\textwidth}
        \centering
        \includegraphics[width=\textwidth]{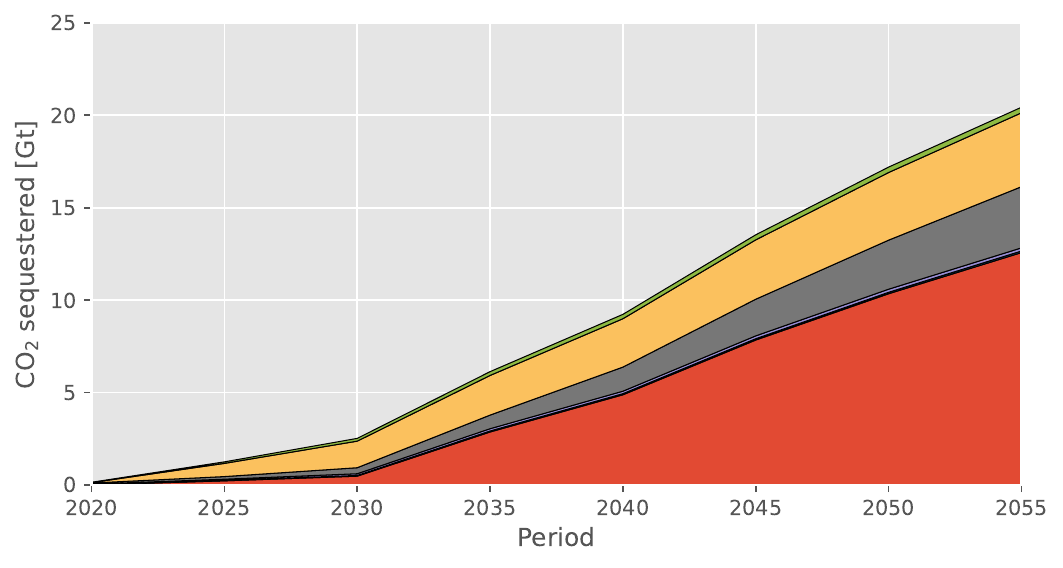}     
        \caption{Affordable, without Russian gas}
        \label{fig:co2_seq_cheap}
    \end{subfigure}
    \begin{subfigure}[htb]{0.49\textwidth}
        \centering
        \includegraphics[width=\textwidth]{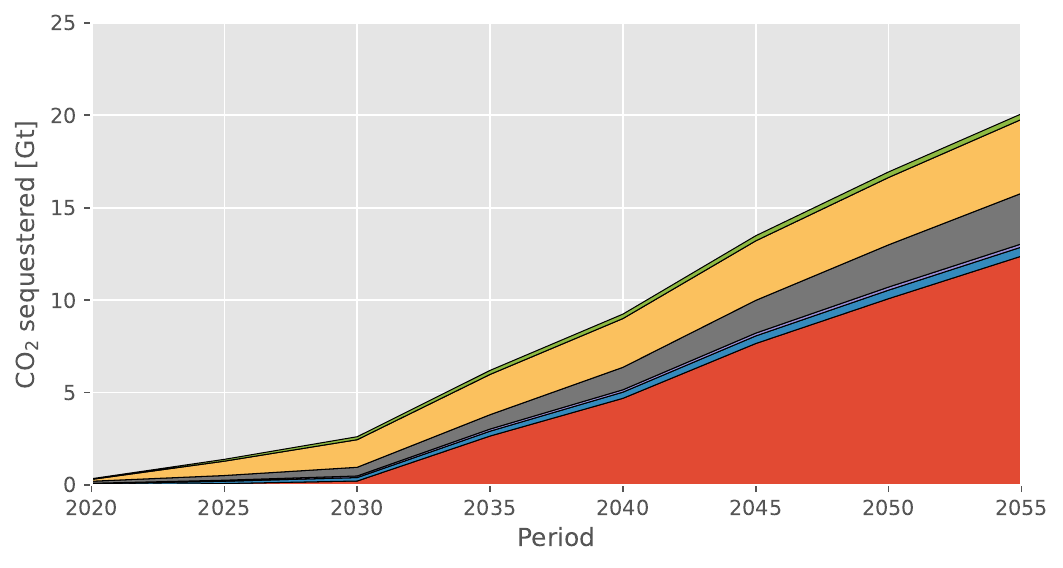}     
        \caption{Costly, without Russian gas}
        \label{fig:co2_seq_exp}
    \end{subfigure}
    \\
    \begin{subfigure}[htb]{0.6\textwidth}
        \centering
        \includegraphics[width=\textwidth]{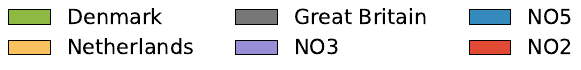}
    \end{subfigure}
    \caption{Expected cumulative amounts of CO$_2$ sequestered in the North Sea.}
    \label{fig:co2_seq}
\end{figure}

The results in this work use CCS on a large scale, and \cref{fig:co2_seq} shows how much CO$_2$ is sequestered in the North Sea until 2055. It is evident that regardless of the case that has been investigated, CO$_2$ sequestration is an effective way to decarbonize the European energy system, and by 2050, at least 10~Gt of CO$_2$ has been sequestered in the North Sea.

Inspecting where the CO$_2$ is sequestered, it becomes clear that the geographic location of the sequestering site is important. The first areas to begin sequestering CO$_2$ are Denmark and the Netherlands, and these are also the only areas to fully utilize their maximum sequestration capacity. Following these two locations, the rest of the captured CO$_2$ is mainly stored in South-Western Norway, NO2, and Great Britain, owing to their proximity to continental Europe.

In the cases without Russian gas, shown in \cref{fig:co2_seq_cheap,fig:co2_seq_exp}, CO$_2$ sequestration is used at a bigger scale than the cases with Russian gas, and at least 20~Gt of CO$_2$ is sequestered in these two cases. In \cref{fig:power} it was shown that without Russian gas, the European energy system would rely more heavily on coal power plants with CCS, which capture more CO$_2$ per unit of energy than their gas-based counterparts. It was also shown in \cref{fig:steel} how CCS played a large role in the steel sector once Russian gas is unavailable, and the effect of these changes is that more CO$_2$ has to be sequestered in the North Sea, as shown in \cref{fig:co2_seq}.

\begin{figure}[ht!]
    \centering
    \begin{subfigure}[htb]{0.4\textwidth}
        \centering
        \includegraphics[width=\textwidth]{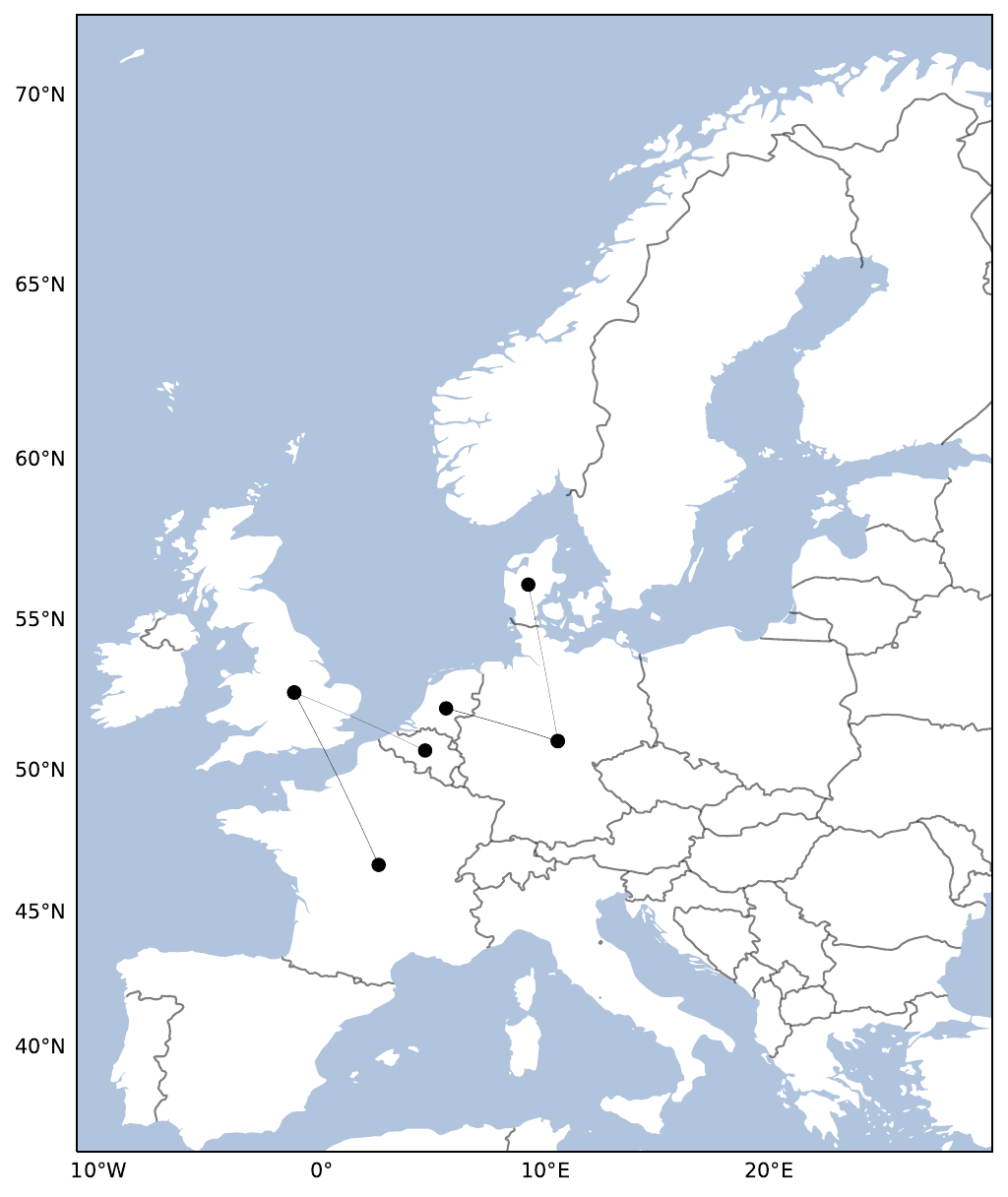}     
        \caption{2030, with Russian gas}
        \label{fig:co2_map_ru_2030}
    \end{subfigure}
    \begin{subfigure}[htb]{0.4\textwidth}
        \centering
        \includegraphics[width=\textwidth]{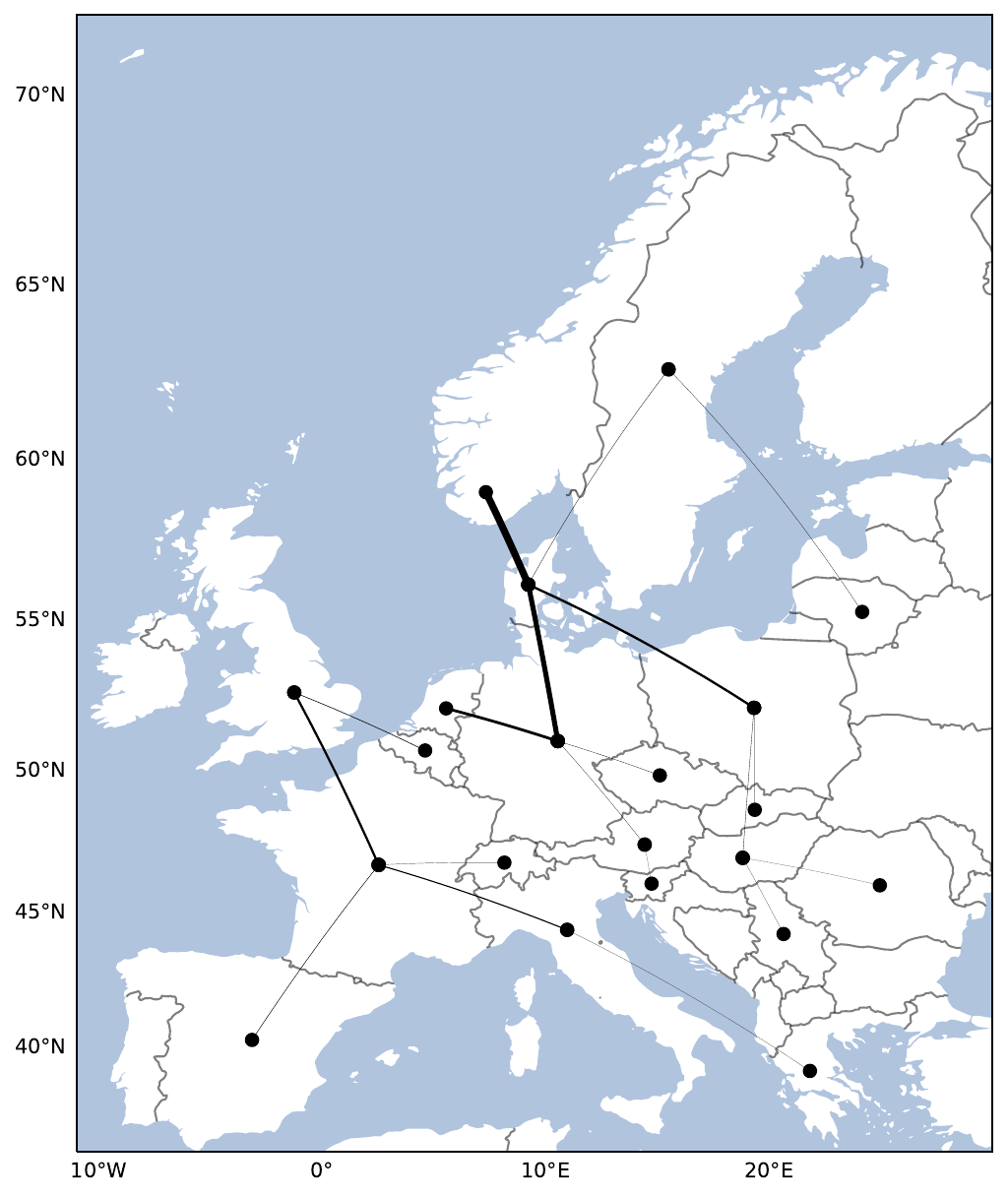}
        \caption{2050, with Russian gas}
        \label{fig:co2_map_ru_2050}
    \end{subfigure}
    \\
    \begin{subfigure}[htb]{0.4\textwidth}
        \centering
        \includegraphics[width=\textwidth]{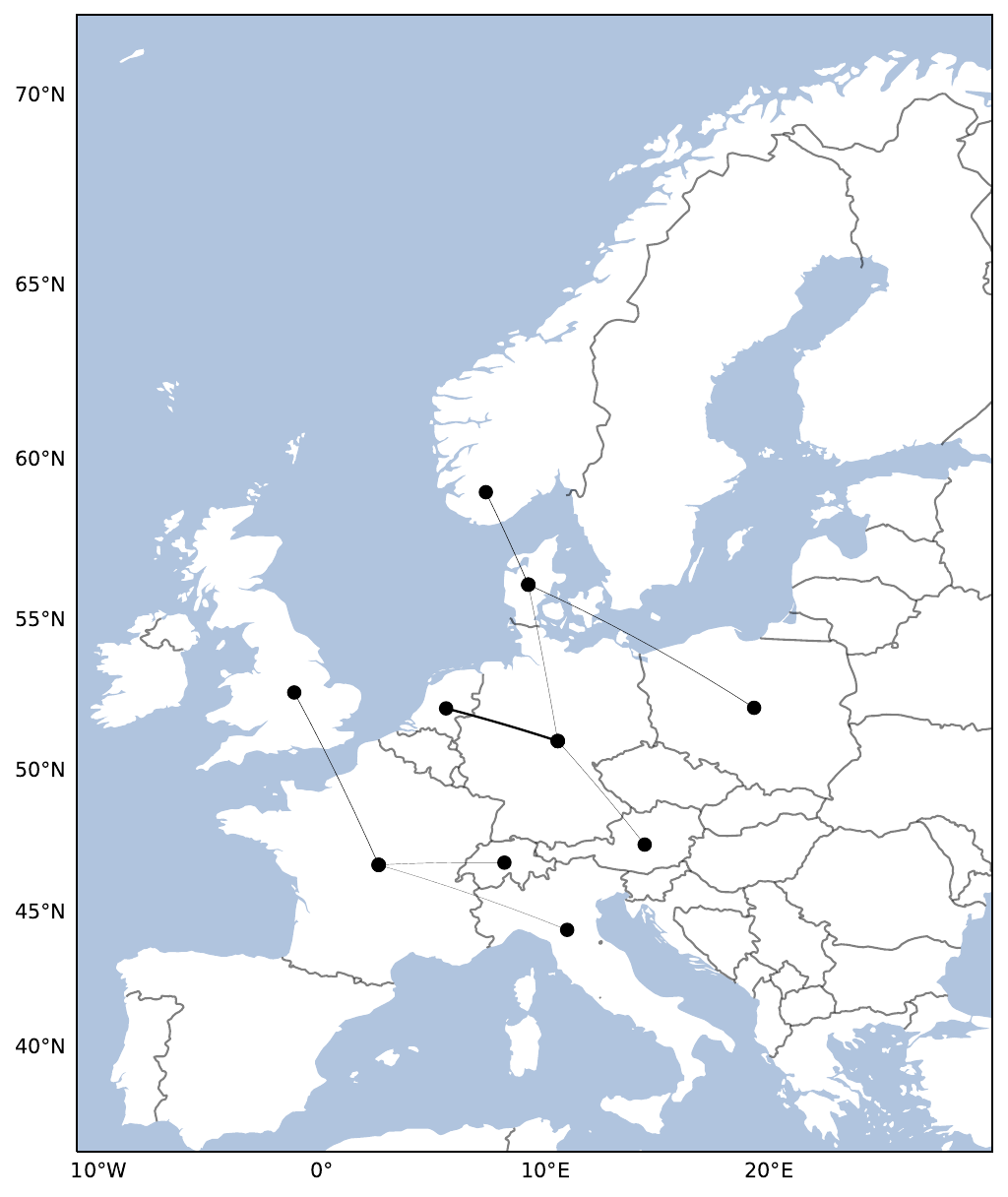}     
        \caption{2030, without Russian gas}
        \label{fig:co2_map_2030}
    \end{subfigure}
    \begin{subfigure}[htb]{0.4\textwidth}
        \centering
        \includegraphics[width=\textwidth]{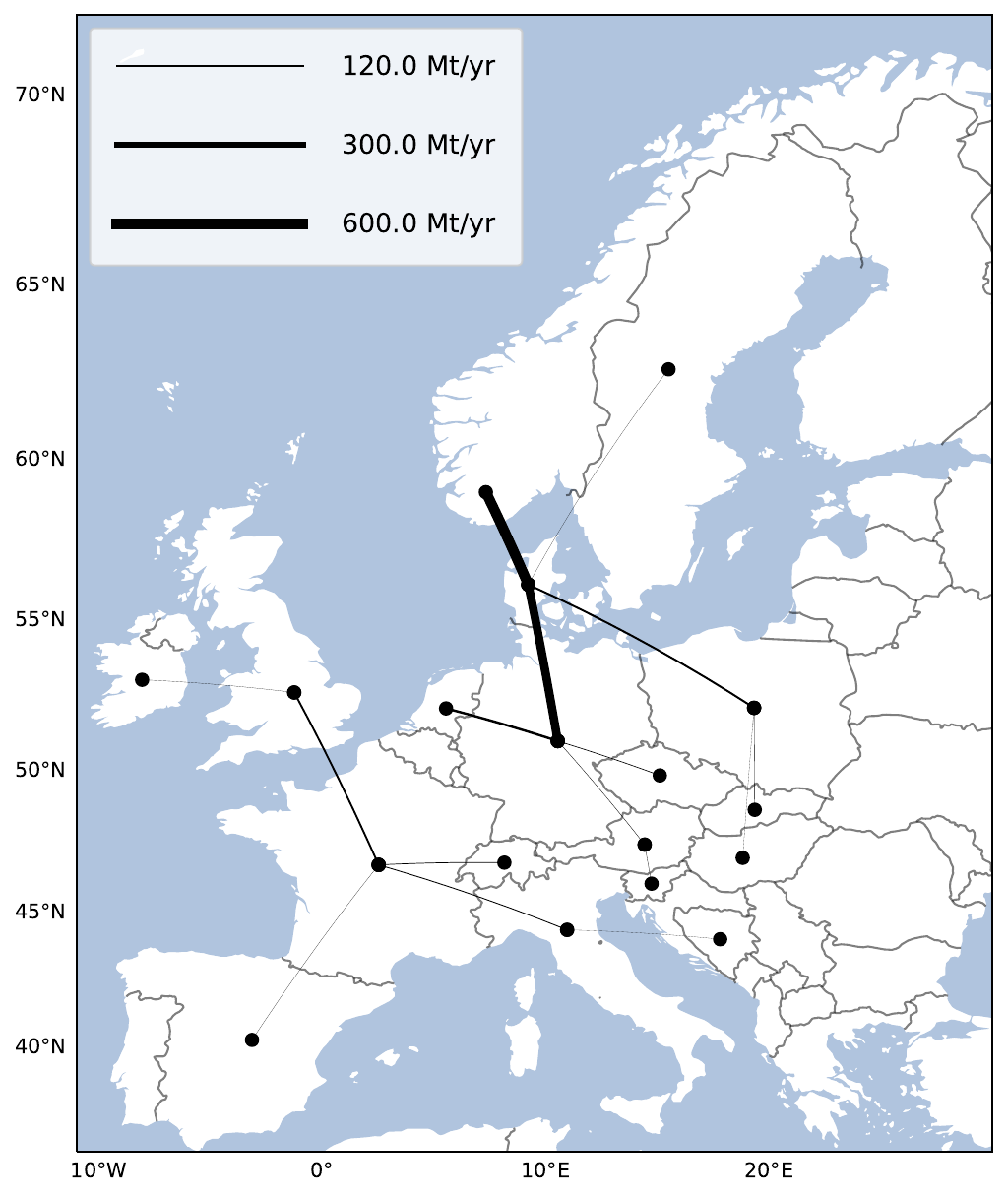}     
        \caption{2050, without Russian gas}
        \label{fig:co2_map_2050}
    \end{subfigure}
    \caption{Development of CO$_2$ pipeline topography. All figures in the costly natural gas case.}
    \label{fig:co2_map}
\end{figure}

\cref{fig:co2_map} shows the  CO$_2$ pipeline topographies in 2030 and 2050 for the costly natural gas cases, with and without Russian gas. Broadly speaking, the topographies in 2050 look very similar for the cases with and without Russian gas, shown in \cref{fig:co2_map_ru_2050,fig:co2_map_2050}. Here, the European countries are well-connected to each other, and with end-points in the main sequestration nodes, showing the importance of CCS in the future.

In 2030, some differences arise. Comparing \cref{fig:co2_map_ru_2030,fig:co2_map_2030}, it can be seen how both cases show the start of the CO$_2$ pipeline networks seen in 2050, but also how the case without Russian gas, shown in \cref{fig:co2_map_2030}, has a much more developed CO$_2$ pipeline network than the case with Russian gas. In fact, the sum of CO$_2$ pipeline capacities in 2030 in the case without Russian gas is over 3 times as large as the case with Russian gas. Furthermore, it is also evident how more countries have adopted CCS by 2030 in \cref{fig:co2_map_2030}, and the topography is consequently more branched out.

CCS was predicted to be an important technology for the industry and power sectors~\cite{Holz2021ASector}, and it appears that it has become even more important following the disconnection from Russian gas. This applies both in the short term, as seen in the 2030 maps in \cref{fig:co2_map}, but also in the long term, as demonstrated in \cref{fig:co2_seq}, where the total CO$_2$ sequestered by 2050 when there is no Russian gas significantly exceeds the cases when Russian gas is available.

\section{Conclusion} \label{sec:conclusion}

This work has investigated how the European energy system can reach the carbon neutrality targets by 2050 in the power, domestic low temperature heat, and industry sectors, while also accounting for the energy demand in the transport sector. The paper has analyzed energy transition pathways without using Russian pipeline gas, and the results were compared with the case where Russian gas would again be available. An important contribution of the work is  endogenous hydrogen demand modelling, enabling the model to optimize the deployment of technologies using hydrogen in the power and industry sectors, taking into account the scarcity of electricity and natural gas, which are required to produce hydrogen.

As a general conclusion from the results, hydrogen is projected  a key role in the industry sectors going forward, and a minor role in the power system. The results show that hydrogen may also play an important role in the domestic heat sector, where it is used as a clean fuel for district heat networks.

The results also show a tremendous value of CCS in the decarbonized European energy system, especially now that Russian pipeline gas is not going to be used. With less affordable natural gas available, the European energy system relies more heavily on coal than it otherwise would, especially in the power and steel industries. This coal use is combined with CCS in order to significantly lower the CO$_2$ emissions.

Summarizing the findings in key messages, it is found that:

\begin{itemize}
    \item \textbf{The removal of Russian natural gas increases the use of coal.} It is found that in the power sector, coal power plants replace the role that gas otherwise would have as a dispatchable generator. In the steel sector, the use of iron reduced using hydrogen is also significantly delayed when Russian gas is unavailable, as the volume of affordable hydrogen in the energy system is insufficient. Consequently, BF-BOF steel plants fuelled by coal are used for longer. In both the power and steel sectors, CCS is used in order to decarbonize coal use.
    \item \textbf{The use of gas in the power sector is partially replaced by renewable power generators.} As access to natural gas becomes more restricted, by first removing access to Russian pipeline gas, and later increasing the price of LNG, it is shown how the generation capacities for the renewables grow considerably. In 2050, wind and solar account for most of the power generation capacity in all cases, but they play a much larger role when LNG is costly and Russian gas is unavailable.
    \item \textbf{Blue hydrogen production is a cost-effective way of producing low-carbon and affordable hydrogen.} Natural gas reforming, both with and without CCS, accounts for a large share of hydrogen production in all investigated cases, and in most cases it is the only source of hydrogen before 2050. Only when Russian gas is unavailable and LNG is very expensive does green hydrogen account for over half of the production capacity in 2050. 
    \item \textbf{CCS is important for reaching European decarbonization goals.} In all the investigated cases in this work, CCS plays a significant role in reducing European greenhouse gas emissions. This is especially the case in the power, hydrogen, and cement sectors. By 2050, at least 10~Gt of CO$_2$ is sequestered in the North Sea in all cases, with Great Britain, the Netherlands and South-Western Norway sequestering the most, owing to their geographic location and maximum offshore sequestration capacity.
    \item \textbf{Phasing out Russian pipeline gas increases the importance of CCS.} In the cases where Russian gas is removed, the minimum amount of CO$_2$ sequestered by 2050 increases to 15~Gt. Furthermore, it is shown that the European CO$_2$ pipeline transport chain develops faster when Russian gas is unavailable. This is a result of how CCS is picked up in the steel industry, and also due to its use with more carbon-intense coal plants in the power sector.
\end{itemize}

There are several ways in which this work can be expanded and improved upon. These include:

\begin{itemize}
    \item \textbf{Including endogenous handling of the transport sector.} This work has an exogenous transport demand for different energy carriers, including hydrogen, natural gas and oil. However, in following with the goal of the work to study the optimal uptake of different low-carbon energy carriers and fuels under different energy market conditions, it would also be worthwhile to treat the transport sector similarly to the other included sectors in this work.
    \item \textbf{Including additional industrial sectors in the model.} This work only includes four industries in the representation of the industry sector: steel, cement, ammonia and oil refining. There are other sectors that are also energy-intensive that are also covered by the ETS, \eg, the aluminium sector. It would be interesting to also include these sectors in this work, to have a more complete representation of European industry.
    \item \textbf{Including long-term uncertainty.} Studying the European energy system until 2050 includes many uncertainties, especially long-term uncertainties when it comes to technology development and future policy. These uncertainties are undoubtedly important to planners today and in the future, and frameworks that include these uncertainties in their planning will be highly valuable. Future works should therefore look for ways in which these can be included while retaining the computational tractability of these problems.
    \item \textbf{Conducting a sensitivity on CCS parameters.} The results in this paper rely heavily on CCS, in all of the sectors that include this technology. However, CCS is not a mature technology yet. It would therefore be valuable to inspect how resilient this pathway is to alternative technological and economical developments in the CCS space. Moreover, studying different policies with regards to CCS acceptance would also be interesting.
\end{itemize}

\FloatBarrier
\section*{Acknowledgments}
This publication has been partially funded by the CleanExport project - Planning Clean Energy Export from Norway to Europe, 308811. The authors gratefully acknowledge the financial support from the Research Council of Norway and the user partners Å Energi, Air Liquide, Equinor Energy, Gassco and TotalEnergies OneTech. The publication has also been partially funded by the Research Council of Norway through the PETROSENTER LowEmission (project code 296207). The authors thank Dr. Julian Straus for valuable inputs to the manuscript. 

\newpage
\appendix

\section{Nomenclature}\label{app:nomenclature}

\begingroup
\begin{multicols}{2}
    \fontsize{11pt}{11pt}\selectfont
    \printnomenclature
\end{multicols}
\endgroup

\section{Introduction to EMPIRE} \label{app:EMPIRE}

This appendix gives an introduction to the structure of EMPIRE, showing the logic of the constraints in the model. For an overview of symbols used in this appendix and their meaning, see \cref{app:nomenclature}.

\cref{eqn:flow_balance_constraint} shows the general formulation of the flow balance for a commodity, $c$, in EMPIRE. The commodities covered by the flow balance constraints include the power, hydrogen, natural gas, CCS, transport, steel, ammonia, cement, and refinery sectors. 

The flow balance consists of sources of a commodity, $\flowSource$, which are the various way in which the commodity is produced. For the power sector for example, the sources are the power generators, and for the natural gas sector, the sources include the various ways of producing or importing natural gas. 

The sinks, $\flowSink$, in the flow balance, are the endogenous uses of the commodity, and this links the different flow balances together. For example, to produce hydrogen with electrolyzers, which is a source in the hydrogen flow balance, it is necessary to consume power, which is a sink in the power flow balance. 

It is also possible to transport some commodities between nodes, which are covered by the two transport variables for import, $\flowIn$, and export, $\flowOut$. Some commodities, such as power, or the cases with inflexible industry, also have exogenous hourly commodity demands that must be met, represented by $\commodityDemand$. Where there is no such hourly demand, $\commodityDemand$ is set to 0. Finally, the power sector uniquely also has the option to curtail demand, which is covered by the variable $\lostLoad$.

\begin{multline}
    \sum_{p\in \mathcal{P}^c} \flowSource - \sum_{sink\in \sigma^c} \flowSink - \sum_{m \in \mathcal{L}^c_n} \left(\flowOut - \flowIn \right) \\= \commodityDemand \ (- \ \lostLoad) 
    \quad \forall n\in \mathcal{N}, h\in \mathcal{H}, i\in \mathcal{I}, w\in {\Omega}
    \label{eqn:flow_balance_constraint}
\end{multline}

\cref{eqn:lifetime_constraint} describes how for an asset $a$, the individual investments into capacity for that asset, $\investmentVariable$ and the remaining initial capacity of that asset, $\remainingInitCap$, sum up to the total capacity of that asset, $\capacityVariable$.

\begin{equation}
    \sum_{j=i'}^{i} \investmentVariable + \remainingInitCap = \capacityVariable
    \quad \forall n\in \mathcal{N}, i\in \mathcal{I}, i'= \max\{1, i-i^{life}_a\}, a \in \mathcal{A}
    \label{eqn:lifetime_constraint}
\end{equation}

An asset cannot be operated, $\operationsAsset$, at a higher level than its capacity, $\capacityVariable$,as described in \cref{eqn:capacity_constraint}.

\begin{equation}
    \operationsAsset \leq \capacityVariable \quad \forall a\in \mathcal{A}, n\in \mathcal{N}, i\in \mathcal{I}, h\in \mathcal{H}, \omega\in \Omega
    \label{eqn:capacity_constraint}
\end{equation}

\cref{eqn:storage_balance_constraint} describes how storage is balanced for the commodities that have storage. In all hours except the first hour of each season, the storage balance simply says that the amount stored at the end of the hour, $\storageOperational$, is the sum of the amount stored in the previous hour, $\storageOperationalPrev$, plus the amount used to charge the storage in this hour, $\storageCharge$, minus the amount discharged from the storage in this hour, $\storageDischarge$.

For those hours that are at the start of a season, a starting amount stored is assumed. In this work, it is assumed that the storage starts half-full, $0.5 \times \capacityStorage$. This is to allow enough flexibility for the model to charge and discharge the storage as it wishes, even during the start of the season.

\begin{subequations}
\begin{align}
     \storageOperationalPrev + \storageCharge - \storageDischarge = \storageOperational &
    \\ \nonumber
    \quad \forall n \in \mathcal{N}, h \in \mathcal{H} \setminus \mathcal{H}^F, i \in \mathcal{I}, \omega \in \Omega \\
    0.5 \times \capacityStorage + \storageCharge - \storageDischarge = \storageOperational &
    \\ \nonumber
    \quad \forall n \in \mathcal{N}, h \in \mathcal{H}^F, i \in \mathcal{I}, \omega \in \Omega 
\end{align}
\label{eqn:storage_balance_constraint}
\end{subequations}

EMPIRE also features a constraint that ensures that the storage level at the last hour of the season is the same as in the start, to ensure that the storage does not lead to a net gain or loss of the commodity in the system. This is shown in \cref{eqn:storage_net_zero_constraint}.

\begin{equation}
   \storageOperational = 0.5 \times \capacityStorage \quad \forall n \in \mathcal{N}, h \in \mathcal{H}^L, i \in \mathcal{I}, \omega \in \Omega
   \label{eqn:storage_net_zero_constraint}
\end{equation}

Some commodities have constraints that apply throughout the entire temporal horizon. This includes the natural gas reserves, where the sum of all natural gas production over all periods cannot exceed the local reserves of natural gas. Similarly, for CCS, it is not possible to sequester more CO$_2$ that the maximum capacity at that geographic location, $\geographicCapacity$. This is described in \cref{eqn:global_constraints}, where the hourly operations are first scaled up to yearly values through the factor $\seasScaleFac$, and then to the length of the period through the factor $\periodlength$. Note that the factor $(\flowSink / \flowSource)$ signifies that either there is a source of the commodity, as with natural gas, or there is a sink of the commodity, as with CO$_2$ in CCS.

\begin{equation}
    \sum_{i \in \mathcal{I}}\sum_{s\in \mathcal{S}}\sum_{h\in \mathcal{H}^s} \periodlength \times \seasScaleFac \times (\flowSink / \flowSource) \leq \geographicCapacity
    \quad \forall n\in \mathcal{N}, \omega \in \Omega
    \label{eqn:global_constraints}
\end{equation}

\bibliographystyle{elsarticle-num-names} 
\bibliography{bibliography}





\end{document}